\documentclass{aims}

\usepackage{amsmath,amssymb,amsthm}
\usepackage{booktabs}
\usepackage{algorithm}
\usepackage{algpseudocode}
\usepackage{adjustbox}
\usepackage{float}
\usepackage{tabularx}
\usepackage{tikz}
\usepackage{xcolor}

\usetikzlibrary{arrows.meta,positioning,shapes,calc}

\allowdisplaybreaks
\usepackage{newtxtext} 
\usepackage{newtxmath} 
\usepackage{cite} 
\emergencystretch=\maxdimen
\usepackage{microtype}

\def\typeofarticle{Research article}
\def\currentvolume{22}
\def\currentissue{5}
\def\currentyear{2026}

\def\ppages{2503--2554}
\def\DOI{10.3934/jimo.2026092}
\def\Received{20 February 2026}
\def\Revised{06 April 2026}
\def\Accepted{20 April 2026}
\def\Published{28 April 2026}

\numberwithin{equation}{section}

\makeatletter
\renewcommand{\@biblabel}[1]{#1\hfill \hspace{-0.2cm}}
\makeatother

\newtheorem{theorem}{Theorem}[section]

\newtheorem{lemma}[theorem]{Lemma}
\newtheorem{proposition}[theorem]{Proposition}

\theoremstyle{definition}

\newtheorem{assumption}[theorem]{Assumption}

\begin{document}

\title{Optimal automation under overdispersed discrete risk: thresholds and hysteresis in a Negative Binomial model}

\author{%
Jinho Cha\affil{1,}\corrauth,
Sahng-Min Han\affil{1},
Long Pham\affil{2},
Joseph Mollick\affil{2}
and Justin Yu\affil{3}
}

\shortauthors{the Author(s)}

\address{%
\addr{\affilnum{1}}{Department of Computer Science, Gwinnett Technical College, Georgia, USA}
\addr{\affilnum{2}}{Department of Decision Sciences and Economics, College of Business, Texas A\&M University--Corpus Christi, Texas, USA}
\addr{\affilnum{3}}{Scheller College of Business, Georgia Institute of Technology, Georgia, USA}
}

\corraddr{Email: \href{mailto:jcha@gwinnetttech.edu}{jcha@gwinnetttech.edu}.}

\begin{abstract}
Decision-making under clustered uncertainty requires models that accommodate discrete overdispersed risk and endogenous control mechanisms. Standard approaches often impose equi-dispersion or Gaussian assumptions and treat control as exogenous or binary, limiting structural realism. In this  paper, we developed a stochastic optimization framework in which risk follows a dynamic Negative Binomial process with time-varying success probability, generating persistent overdispersion and temporal dependence. A continuous control variable mitigated non-tail and tail exposure under convex implementation and adjustment costs. Analytical results showed that overdispersion reshapes the optimal policy, generating interior solutions, dispersion-dependent thresholds, and regime-dependent hysteresis effects that do not arise under equi-dispersed specifications. Simulation evidence confirms that overdispersion amplifies tail risk and induces nonlinear responses. An empirical analysis using daily S\&P 500 index data documents significant overdispersion and persistent clustering. The results provide a structural basis for integrating discrete risk modeling with dynamic operational decisions under uncertainty.
\end{abstract}


\maketitle

\section{Introduction}
\label{sec:intro}

Modern financial and operational systems are increasingly exposed to
clustered risk, in which shocks arrive in bursts and volatility regimes
persist over time
\cite{Castellano2024SystemicRisk, Caporin2024EarlyWarnings, Bergmann2026ExtremeRiskClustering}.
Empirical evidence from equity and currency markets shows that shocks are
neither independent nor evenly distributed over time. Realized events
exhibit overdispersion, temporal dependence, and pronounced clustering
\cite{Cont2001StylizedFacts, ContTankov2004Financial}. These features imply
that risk variability is state-dependent and can remain elevated for
extended periods.

In such environments, conventional modeling assumptions are structurally
inadequate. The classical Poisson framework imposes an equi-dispersion
restriction that mechanically ties the variance of event counts to their
mean, an assumption long recognized as empirically fragile in count-data
applications
\cite{Hausman1984CountData, CameronTrivedi1990Overdispersion}. By
construction, this restriction rules out persistent overdispersion and
cannot accommodate regimes in which risk variability systematically exceeds
its average level. Gaussian approximations, while analytically convenient,
abstract from discreteness and are poorly suited to represent clustered
jump arrivals. Financial econometrics has repeatedly documented volatility
clustering, conditional heteroskedasticity, and long-memory dynamics in
asset returns
\cite{Bollerslev1992ARCH, Lo1991LongTermMemory}, indicating that
dispersion dynamics are structural features rather than statistical
anomalies.

When dispersion is underestimated, policy prescriptions become
systematically distorted. Optimal execution strategies, clearing rules, and
risk-management mechanisms derived under homoskedastic Gaussian or
equi-dispersed Poisson assumptions may understate tail exposure and
systemic fragility. In particular, ignoring persistent overdispersion can
bias investment decisions in monitoring, automation, and protective
infrastructure
\cite{Dean1992Overdispersion, Dean2016Overdispersion, Berk2008Overdispersion, Winkelmann2008Poisson}.

\textls[-30]{Automation and smart execution mechanisms have become central to modern
financial infrastructures
\cite{Almgren2001Optimal, Biais2019Blockchain, GoldfarbTucker2019Digital, Schar2021DeFi, Cartea2015Algorithmic}.
Algorithmic trading systems, automated clearing platforms, and rule-based
smart contracts enable rapid state-contingent execution. Furthermore,
automation entails implementation costs, adjustment frictions, and
potential rigidity
\cite{Almgren2001Optimal, Cartea2015Algorithmic}. Greater automation
intensity may mitigate losses and reduce reaction delays, but it also
reduces flexibility and introduces switching costs. This trade-off gives
rise to a structural optimization problem: How should automation intensity
be chosen when risk arrives discretely and clusters persistently over time?}      

To address this question, the framework developed in this paper represents
discrete loss or demand events using a Negative Binomial (NB) count process,
which explicitly accommodates overdispersion and clustering
\cite{hilbe2011negative, Klein2015GAMLSS}. Automation intensity is modeled
as a continuous control variable that mitigates loss exposure while
incurring convex implementation and adjustment costs. The analysis begins
with a static optimization problem and extends to a dynamic
regime-switching environment in which clustered risk evolves persistently
across states. Under this formulation, dispersion is not a second-order
statistical correction but a first-order structural determinant of the
optimal automation architecture.

\medskip
\noindent
\textbf{Contributions.}
The major contributions are as follows:

\begin{itemize}
    \item[(i)] \textls[-25]{A structural optimization framework is established in which
    clustered discrete risk reshapes the geometry of optimal automation
    policies, generating threshold effects absent under equi-dispersed
    models.}

    \item[(ii)] Overdispersion is shown to affect not only expected losses
    but also the structure of policy responses, producing interior
    solutions and regime-dependent automation thresholds.

    \item[(iii)] \textls[-25]{In a dynamic regime-switching environment with adjustment
    frictions, endogenous hysteresis arises, implying that
    crisis-induced automation need not fully revert even when underlying
    conditions normalize.}

    \item[(iv)] Empirical evidence based on S\&P 500 jump measures supports
    these mechanisms by showing that dispersion varies across regimes and
    that the response to dispersion becomes regime-dependent, including
    reversal behavior in high-dispersion states.
\end{itemize}

\medskip
\noindent
\textbf{Empirical motivation.}
Daily extreme-return (jump) counts for the S\&P 500 provide a natural
setting in which clustered dispersion is empirically salient
\cite{BarndorffNielsenShephard2004, AitSahaliaJacod2009}. Volatility spike
counts, such as threshold exceedances of the Chicago Board Options Exchange Volatility Index (CBOE VIX), and insurance claim
frequencies provide additional examples that typically exhibit
overdispersion
\cite{BollerslevTodorov2011, Denuit2006Actuarial}. These series provide
natural calibration environments in which clustered dispersion is
economically meaningful.

The remainder of the paper is organized as follows:
In section~\ref{sec:static}, we present the static formulation, including the
dispersion-based threshold structure.
In Section~\ref{sec:dynamic}, we develop the dynamic extension with regime
switching and adjustment frictions.
In Section~\ref{sec:synthetic_experiments}, we provide synthetic structural experiments.
In Section~\ref{sec:empirical_sp500}, we present the empirical analysis and implementation.
In Section~\ref{sec:conclusion}, we present the structural implications, integrate the theoretical and empirical findings, and conclude.

\section{Literature review}
\label{sec:lit}

\subsection{Clustered risk and overdispersion}
\label{subsec:lit_overdispersion}

Empirical literature documents that risk in financial markets is
not well described by independent and homoskedastic shocks.
Instead, extreme events tend to arrive in clusters,
volatility exhibits persistence,
and tail realizations concentrate in episodic bursts.
Volatility clustering is a foundational stylized fact,
formalized through conditional heteroskedasticity models
such as autoregressive conditional heteroskedasticity (ARCH) and
generalized autoregressive conditional heteroskedasticity (GARCH)
\cite{Engle1982,Bollerslev1986},
while high-frequency evidence shows that jump components
occur in concentrated episodes rather than as isolated arrivals
\cite{BarndorffNielsenShephard2004,AitSahaliaJacod2009}.
Further, tail risk premia vary over time,
indicating regime-dependent compensation for rare events
\cite{BollerslevTodorov2011}.
These findings imply that risk intensity is persistent,
state-dependent, and frequently exhibits dispersion
beyond that implied by equi-dispersed count models.

Within discrete-event modeling,
overdispersion has long been recognized as a central empirical issue.
Early econometric contributions demonstrated that
Poisson specifications are often rejected because
the variance of observed counts exceeds the mean,
motivating the use of NB models
as Poisson–Gamma mixtures that accommodate excess variance
\cite{Hausman1984CountData,CameronTrivedi1990Overdispersion}.
In actuarial science, NB and related mixed-Poisson models
are standard tools for modeling clustered claim arrivals
and heterogeneous risk exposure
\cite{Denuit2006Actuarial}.
The defining property of the NB specification is that
$\mathrm{Var}(D_t)$ exceeds $\mathbb{E}[D_t]$ in a
structurally interpretable manner,
enabling dispersion to serve as a primitive parameter
rather than a residual anomaly.

Despite its widespread use in statistics and actuarial applications,
the NB process has played only a limited role in financial control problems.
In financial econometrics, clustering is typically addressed
through stochastic volatility or jump–diffusion frameworks,
while discrete event counts are rarely embedded directly
into optimization problems.
Similarly, in the optimal execution literature,
endogenous control variables such as trading intensity
are chosen under stochastic cost dynamics,
but incoming risk is usually modeled as Brownian motion
or as a simple Poisson arrival process
\cite{Almgren2001Optimal,Cartea2015Algorithmic}.
These specifications abstract from persistent overdispersion
and therefore do not capture clustered count risk.

To the best of our knowledge,
no researcher has integrated an overdispersed count process
into a structural optimization framework in which
(i) risk arrives discretely,
(ii) dispersion is state-dependent,
(iii) control variables are chosen endogenously,
Researchers treat these elements separately,
but their joint interaction remains unexplored.

This gap is economically nontrivial.
When dispersion varies across regimes,
risk is not merely larger in expectation;
its temporal geometry changes.
Embedding an NB process into a dynamic control problem
therefore alters the shape of optimal policies,
potentially generating threshold effects,
interior adoption regions,
and hysteresis phenomena
that do not arise under equi-dispersed specifications.
In this study, we develop such an integrated framework.

This structural gap becomes particularly salient
in environments where execution mechanisms are automated.
When policies are implemented through event-triggered systems,
the statistical representation of risk directly interacts
with the architecture of execution.

\subsection{Automation and digital execution}
\label{subsec:lit_automation}

A parallel stream of research examines how digitization reshapes the
\emph{execution layer} of contracting. Blockchain-based systems expand
the contracting space through verifiable state-contingent execution,
altering information transmission and equilibrium structure
\cite{CongHe2019Blockchain,ChiuKoeppl2019RFS}.
In financial markets, decentralized consensus and smart contracts
modify how agreements are triggered and enforced,
with implications for competition, transparency,
and strategic interaction
\cite{Biais2019Blockchain,Momtaz2024DeFi,Pantelidis2025Crypto}.

In traditional finance, automation is embodied in
algorithmic execution and automated clearing systems.
The optimal execution literature formalizes how trading algorithms
translate policy variables into executed quantities
while managing market impact and risk exposure
\cite{Almgren2001Optimal,KyleObizhaevaWang2018}.
Subsequent treatments incorporate market microstructure dynamics,
inventory control, and real-time decision rules under uncertainty
\cite{Cartea2015Algorithmic,BiaisFoucaultMoinas2011,
FoucaultKozhanTham2017,EasleyLopezdePradoOHara2012RFS,
Ladley2020SpeedSophistication}.
Across these settings, execution rules convert information into actions
through predefined response mechanisms.

A common structural feature emerges across blockchain protocols
and algorithmic trading systems:
Execution is typically \emph{event-contingent}.
Actions are triggered when discrete conditions are satisfied,
such as order arrivals, margin thresholds, or state transitions,
rather than responding continuously to infinitesimal diffusion shocks.
Smart contracts activate upon verifiable events,
and execution algorithms react to observable order flow
or price-impact signals
\cite{CongHe2019Blockchain,Almgren2001Optimal,
BiaisFoucaultMoinas2011,
DuffieDworczakZhu2017,
FoucaultKozhanTham2017}.

This architecture aligns naturally with discrete,
count-based representations of risk.
When shocks arrive in clustered bursts,
the execution layer absorbs episodic spikes rather than smooth flows.
A model based solely on Gaussian diffusion abstracts from this
event-driven structure. Continuous diffusion implies smooth policy adjustments
to infinitesimal shocks,
whereas event-driven architectures respond to discrete realizations.
Ignoring this distinction conflates volatility with arrival intensity
and may distort optimal adoption boundaries.
By contrast, an overdispersed count process captures
the discreteness of activation triggers
and the clustering of shocks that automated systems must process.

From this perspective, adopting a NB specification
is not merely a statistical device to accommodate excess variance.
It reflects the informational environment faced by automated systems:
Risk materializes through discrete arrivals whose intensity
varies across regimes.
Embedding such a process into a policy-choice framework
therefore mirrors the operational logic of digital execution.

\subsection{Dynamic adjustment and hysteresis}
\label{subsec:lit_dynamic}

When risk arrives in clustered bursts and execution mechanisms are
event-contingent, a purely static adoption decision becomes incomplete.
Automation is not chosen once and for all; rather, it is adjusted over time
in response to evolving risk regimes.
Firms may adopt automation during stress episodes, pause during calm periods,
and delay reversal when uncertainty remains elevated.
Such behavior naturally generates \emph{hysteresis} driven by switching frictions
and regime persistence.

The economics of irreversible or partially irreversible investment provides a
canonical foundation for this phenomenon. With fixed or quasi-fixed adoption
costs, optimal policies exhibit inaction regions and threshold behavior:
Adjustment occurs only when state variables cross critical boundaries,
implying history dependence even when fundamentals mean-revert
\cite{DixitPindyck1994, Caballero1991}. In these models,
volatility reshapes optimal entry and exit regions, leading firms to delay
otherwise positive positive net present value investments. This mechanism provides a natural
microfoundation for dispersion-dependent threshold responses in our model.

Asset-pricing and macro-finance research further emphasizes that
persistent second-moment dynamics alter optimal decision rules.
Long-run risk models show that time-varying volatility and growth
uncertainty materially affect equilibrium investment and pricing
\cite{BansalYaron2004}.
Models with rare disasters and time-varying tail risk demonstrate that
low-probability clustered shocks have first-order effects on equilibrium
behavior and return dynamics
\cite{Barro2006,Gabaix2012,Wachter2013}.
These frameworks imply that when risk intensity varies across regimes,
optimal policies respond nonlinearly and asymmetrically.

Risk-sensitive and robustness-based control approaches provide an additional
mechanism for persistence.
When agents account for model uncertainty or tail ambiguity,
optimal thresholds shift across regimes and reversals are delayed,
amplifying hysteresis under clustered uncertainty
\cite{HansenSargent2001,AndersonHansenSargent2003,BarillasHansenSargent2009}.
In macro-financial data, downside risk and growth vulnerability
vary systematically with financial conditions,
consistent with regime-dependent amplification
\cite{AdrianBoyarchenkoGiannone2019}.

\textls[-25]{A complementary approach formalizes clustered risk through
regime-switching dynamics.
Markov regime-switching models parsimoniously capture
persistent states such as “normal” and “stress” environments
\cite{Hamilton1989,AngBekaert2002}.
Embedding such regimes into dynamic control problems yields tractable
policy structures in which adjustment costs and state persistence
jointly generate threshold behavior and delayed responses.}

These strands of literature motivate three structural ingredients.
First, an explicit state space $\mathcal{S}$ captures persistent regimes.
Second, dispersion measures such as $\mathrm{Var}(D_t)$ summarize
time-varying overdispersion in clustered count risk.
Third, a dynamic automation variable evolves subject to switching costs.
Together, these components support a transparent
``static core + dynamic option'' architecture:
The static model characterizes equilibrium adoption under given dispersion,
while the dynamic extension explains asymmetric entry–exit bands
and persistent automation regimes.
In particular, clustered overdispersion shifts optimal adoption thresholds
endogenously, generating hysteresis even when average risk intensity
normalizes.

The preceding strands of literature establish three structural facts:
(i) Clustered risk generates state-dependent overdispersion,
(ii) automated execution operates through event-contingent mechanisms, and
(iii) irreversible adjustment under persistent regimes produces hysteresis.
What remains missing is a unified optimization framework that integrates
these elements into a single tractable structure.

In the next section, we develop such a framework.
We first construct a static automation problem in which clustered risk
enters exclusively through dispersion-sensitive linear operators on the
probability simplex.
This formulation isolates the geometric effect of overdispersion on optimal
automation intensity while holding the mean risk level fixed.
We then embed the static core into a dynamic regime-switching model with
adjustment frictions, allowing dispersion to evolve endogenously across
persistent states and generating structural hysteresis.

\section{Static model under clustered risk}
\label{sec:static}

\subsection{Probability-space primitives and operator representation}
\label{subsec:static_prob_operator}

We begin with the exact NB model on the full support
$\mathbb{N}_0$, which provides the theoretical basis for the dispersion
comparative statics developed below. For numerical implementation,
we introduce a truncated finite-dimensional representation
\[
\mathcal{D}=\{0,1,\dots,M\},
\]
and represent the clustered-risk count random variable through a probability
vector $\boldsymbol{\pi}(r)\in\Delta^{M+1}$ induced by the truncated and
renormalized NB distribution.

To isolate pure dispersion effects, we fix the mean $\mu>0$ and parameterize
the NB distribution using the mean-preserving form
\[
D_r \sim \mathrm{NB}\bigl(r,\theta(r)\bigr),
\qquad
\theta(r)=\frac{r}{r+\mu},
\]
so that
\[
\mathbb{E}[D_r]=\mu,
\qquad
\mathrm{Var}(D_r)=\mu+\frac{\mu^2}{r}.
\]
Smaller $r$ implies stronger overdispersion and heavier tail clustering.

Let $p_d(r)=\mathbb{P}(D_r=d)$ for $d\in\mathbb{N}_0$ denote the full NB pmf.
Define the truncated and renormalized probabilities on $\mathcal{D}$ by
\[
\pi_d(r)
=
\frac{p_d(r)}{\sum_{k=0}^{M} p_k(r)},
\qquad d=0,\dots,M,
\]
and collect them as
\[
\boldsymbol{\pi}(r)
=
\big(\pi_0(r),\dots,\pi_M(r)\big)^\top,
\qquad
\mathbf{1}^\top\boldsymbol{\pi}(r)=1,
\]
so that $\boldsymbol{\pi}(r)\in\Delta^{M+1}$.

For any vector $\boldsymbol{g}=(g(0),\dots,g(M))^\top$,
\begin{equation}
\mathbb{E}_{\boldsymbol{\pi}(r)}[g(D_r)]
:=
\sum_{d=0}^M g(d)\pi_d(r)
=
\boldsymbol{g}^\top \boldsymbol{\pi}(r).
\label{eq:linexp}
\end{equation}

Let $\boldsymbol{\ell}=(\ell(0),\dots,\ell(M))^\top$ with $\ell(d)\ge0$ and
nondecreasing in $d$. Fix a stress threshold $d^*\in\mathcal{D}$ and define the
tail-selection operator
\[
\mathsf{T}
:=
\mathrm{diag}\!\big(\mathbf{1}\{d\ge d^*\}\big)
\in\mathbb{R}^{(M+1)\times(M+1)}.
\]
Define the non-tail and tail exposure components by
\begin{equation}
B_1(r):=\boldsymbol{\ell}^\top(\mathsf{I}-\mathsf{T})\boldsymbol{\pi}(r),
\qquad
B_2(r):=\boldsymbol{\ell}^\top\mathsf{T}\boldsymbol{\pi}(r).
\label{eq:B1B2_def}
\end{equation}
Equivalently, the tail functional is
\begin{equation}
\mathrm{Tail}(\boldsymbol{\ell};\boldsymbol{\pi}(r))
=
B_2(r)
=
(\mathsf{T}\boldsymbol{\ell})^\top\boldsymbol{\pi}(r).
\label{eq:tail_linear}
\end{equation}

\subsection{Tail monotonicity under mean-preserving dispersion}
\label{subsec:NB_tail_lemma}

\begin{lemma}[NB tail monotonicity under mean preservation]
\label{lem:NB_tail_monotonicity}
Let $D_r \sim \mathrm{NB}(r,\theta(r))$ with
\[
\theta(r)=\frac{r}{r+\mu}, \qquad \mu>0,
\]
defined on $\mathbb{N}_0$ (no truncation). Define the tail loss function
\[
h(d):=\ell(d)\mathbf{1}\{d \ge d^*\},
\]
and the tail exposure
\[
B_2(r):=\mathbb{E}[h(D_r)]
=
\sum_{d=d^*}^{\infty} \ell(d)\,\mathbb{P}(D_r=d).
\]
Assume that the function $h : \mathbb{N}_0 \to \mathbb{R}_+$ is nonnegative
and nondecreasing, and is convex in the discrete sense, meaning that its
second-order forward differences satisfy
\[
\Delta^2 h(d)
:=
h(d+2) - 2h(d+1) + h(d)
\ge 0
\quad \text{for all } d \in \mathbb{N}_0.
\]

Then $B_2(r)$ is decreasing in $r$.  
If $h$ is strictly convex on a nontrivial subset of $\mathbb{N}_0$
that is assigned positive probability under the induced Poisson mixtures,
then $B_2(r)$ is strictly decreasing in $r$.
\end{lemma}

\begin{proof}
Under the mean-preserving parameterization, the NB admits the
Gamma--Poisson mixture representation:
\[
D_r \mid \Lambda_r \sim \mathrm{Poisson}(\Lambda_r),
\qquad
\Lambda_r \sim \mathrm{Gamma}(r,\mu/r),
\]
with
\[
\mathbb{E}[\Lambda_r]=\mu,
\qquad
\mathrm{Var}(\Lambda_r)=\mu^2/r.
\]

Define
\[
g(\lambda)
:=
\mathbb{E}[h(N_\lambda)],
\qquad
N_\lambda \sim \mathrm{Poisson}(\lambda).
\]
Thus, $g(\lambda)$ represents the expected tail loss generated by
a Poisson count process with intensity $\lambda$.
Then, by the law of total expectation,
\[
B_2(r)=\mathbb{E}[g(\Lambda_r)].
\]

\paragraph{Step 1: Monotonicity of $g$.}
Because the Poisson distribution is increasing in $\lambda$ in the sense of
first-order stochastic dominance, and $h$ is nondecreasing, it follows that
$g(\lambda)$ is nondecreasing in $\lambda$.

\paragraph{Step 2: Convexity of $g$.}
For Poisson expectations, the discrete derivative identities hold:
\[
g'(\lambda)=\mathbb{E}[\Delta h(N_\lambda)],
\qquad
g''(\lambda)=\mathbb{E}[\Delta^2 h(N_\lambda)],
\]
where
\[
\Delta h(d)=h(d+1)-h(d),
\qquad
\Delta^2 h(d)=h(d+2)-2h(d+1)+h(d).
\]
Since $h$ is nondecreasing, $\Delta h(d)\ge0$, hence $g'(\lambda)\ge0$.
Since $h$ is convex, $\Delta^2 h(d)\ge0$, hence $g''(\lambda)\ge0$.
Thus $g$ is increasing and convex.

\paragraph{Step 3: Dispersion ordering of $\Lambda_r$.}
For fixed mean $\mu$, the Gamma family $\mathrm{Gamma}(r,\mu/r)$ becomes
more dispersed as the shape parameter $r$ decreases while preserving the same mean.
Equivalently, if $r_1 > r_2$, then $\Lambda_{r_2}$ dominates $\Lambda_{r_1}$
in convex order (i.e., $\Lambda_{r_2} \succeq_{cx} \Lambda_{r_1}$).
This is a standard ordering property of the Gamma family under
mean-preserving scale adjustment (see, e.g., \cite{ShakedShanthikumar2007_Ch1}).

\paragraph{Step 4: Comparison of expectations.}
Since $g$ is convex,
\[
\mathbb{E}[g(\Lambda_{r_1})]
\le
\mathbb{E}[g(\Lambda_{r_2})]
\quad \text{whenever } r_1 > r_2.
\]
Therefore $B_2(r)$ is decreasing in $r$.

Strict monotonicity follows when $g$ is strictly convex on a region
that is visited with positive probability under the induced Poisson mixtures,
which holds if $h$ is strictly convex on a nontrivial subset of $\mathbb{N}_0$.
\end{proof}

The monotonicity result in Lemma~\ref{lem:NB_tail_monotonicity} is stated
for the exact untruncated NB model on $\mathbb{N}_0$.
The truncated representation introduced above is used only for
finite-dimensional computation and numerical implementation.
For sufficiently large truncation level $M$, it provides an accurate
approximation to the exact tail exposure while preserving the same comparative-static interpretation of dispersion.
    
\subsection{Automation intensity, mitigation operators, and cost mapping}
\label{subsec:static_automation}

The firm chooses an automation intensity $a\in[0,1]$, interpreted as the degree
of automated execution (e.g., algorithmic clearing, smart-contract coverage,
or rule-based enforcement), which is applied to mitigate clustered loss exposure.


Let $\mathsf{I}\in\mathbb{R}^{(M+1)\times(M+1)}$ denote the identity operator
and recall the tail-selection operator 
$\mathsf{T}\in\mathbb{R}^{(M+1)\times(M+1)}$ defined in
\eqref{eq:tail_linear}.  
We have the orthogonal decomposition

\[
\mathsf{I}
=
(\mathsf{I}-\mathsf{T})
+
\mathsf{T}.
\]

Hence total expected loss admits the decomposition

\[
\boldsymbol{\ell}^\top \boldsymbol{\pi}(r)
=
\boldsymbol{\ell}^\top (\mathsf{I}-\mathsf{T}) \boldsymbol{\pi}(r)
+
\boldsymbol{\ell}^\top \mathsf{T}\boldsymbol{\pi}(r).
\]

This guarantees that tail exposure is isolated without double-counting
and provides a clean decomposition of total expected loss.

Automation may affect non-tail and tail exposure differently.
Let

\[
\phi,\psi : [0,1]\to[0,1]
\]

be twice continuously differentiable functions, satisfying

\[
\phi(0)=\psi(0)=0,
\qquad
\phi'(a)\ge 0,
\qquad
\psi'(a)\ge 0,
\]
\[
\phi''(a)\le 0,
\qquad
\psi''(a)\le 0.
\]

Here:

- $\phi(a)$ reduces non-tail exposure,
- $\psi(a)$ reduces tail exposure.

Enabling $\psi\neq\phi$ captures disproportionate mitigation of extreme states.


Expected loss under dispersion parameter $r$ is

\begin{equation}
\begin{aligned}
L(a;r)
&=
(1-\phi(a))
\,\boldsymbol{\ell}^\top (\mathsf{I}-\mathsf{T})\boldsymbol{\pi}(r)
\\
&\quad+
(1-\psi(a))
\,\boldsymbol{\ell}^\top \mathsf{T}\boldsymbol{\pi}(r).
\end{aligned}
\label{eq:expected_loss_operator}
\end{equation}

Using the decomposition in \eqref{eq:B1B2_def}, expected loss under automation becomes

\begin{equation}
L(a;r)
=
B_1(r)\bigl(1-\phi(a)\bigr)
+
B_2(r)\bigl(1-\psi(a)\bigr).
\label{eq:loss_scalar_form}
\end{equation}

Observe that

\[
B_1(r)+B_2(r)
=
\boldsymbol{\ell}^\top \boldsymbol{\pi}(r).
\]

Thus, $B_1(r)$ captures non-tail exposure and $B_2(r)$ captures tail exposure.
Here, $B_1(r)$ and $B_2(r)$ denote the truncated finite-dimensional
approximations of the corresponding exposure components defined in
Section~\ref{subsec:static_prob_operator}.


Automation incurs a convex implementation cost

\begin{equation}
K:[0,1]\to\mathbb{R}_+,
\qquad
K'(a)\ge 0,
\qquad
K''(a)\ge 0.
\label{eq:static_cost_K}
\end{equation}

A canonical quadratic specification is

\[
K(a)=k_1 a + k_2 a^2,
\qquad k_2>0.
\]


Total static expected cost is

\begin{equation}
C(a;r)
=
L(a;r)
+
K(a),
\label{eq:static_cost}
\end{equation}

i.e.,

\begin{equation}
C(a;r)
=
B_1(r)\bigl(1-\phi(a)\bigr)
+
B_2(r)\bigl(1-\psi(a)\bigr)
+
K(a).
\label{eq:static_cost_explicit}
\end{equation}

The firm solves

\begin{equation}
\min_{a\in[0,1]} C(a;r).
\label{eq:static_problem}
\end{equation}

In many financial environments, automation is implemented through
state-contingent mechanisms that activate under stress (e.g., margin calls,
circuit breakers, and   liquidation triggers), implying that $\psi'(a)$ may exceed
$\phi'(a)$ over relevant regions.

\subsection{Structural assumptions}
\label{subsec:static_assumptions}

We now state the structural assumptions required for
well-posedness, existence, uniqueness, and dispersion-based
comparative statics of the static automation problem.

\begin{assumption}[Dispersion and exposure regularity]
\label{ass:dispersion}
Let $\boldsymbol{\pi}(r)\in\Delta^{M+1}$ denote the truncated and renormalized
NB probability vector introduced in
Section~\ref{subsec:static_prob_operator}. The normalization constant $Z(r)=\sum_{k=0}^M p_k(r)$ is strictly positive
and continuously differentiable in $r$, ensuring that $\boldsymbol{\pi}(r)$
is continuously differentiable. Then:

\begin{enumerate}
    \item[(i)] The exposure components
    \[
    B_1(r)
    :=
    \boldsymbol{\ell}^\top (\mathsf{I}-\mathsf{T}) \boldsymbol{\pi}(r),
    \qquad
    B_2(r)
    :=
    \boldsymbol{\ell}^\top \mathsf{T}\boldsymbol{\pi}(r),
    \]
    are continuously differentiable in $r$.

    \item[(ii)] The loss function $\ell(d)$ is nonnegative
    and nondecreasing in $d$.
\end{enumerate}

Under this parameterization, smaller values of $r$
correspond to stronger overdispersion.
Monotonicity of the exact tail exposure with respect to dispersion
is established in Lemma~\ref{lem:NB_tail_monotonicity}.
The truncated representation used here provides a finite-dimensional
approximation that preserves the same comparative-static behavior
for sufficiently large truncation level $M$.
\end{assumption}

\begin{assumption}[Loss boundedness]
\label{ass:loss}
The loss function $\ell:\mathbb{N}_0\to\mathbb{R}_+$ is nonnegative and
nondecreasing, that is, $\ell(d)\ge 0$ for all $d\in\mathbb{N}_0$
and $\ell(d+1)\ge \ell(d)$ for all $d$.
Moreover, it is uniformly bounded in the sense that
\[
\sup_{d\ge 0} \ell(d) < \infty.
\]
These conditions ensure finite expectations and preserve
monotonicity of tail exposure.
\end{assumption}

\begin{assumption}[Automation effectiveness]
\label{ass:automation_effectiveness}
The mitigation functions $\phi,\psi:[0,1]\to[0,1]$
represent the proportion of reduction in non-tail and tail exposure,
respectively. They are twice continuously differentiable and satisfy
\[
\phi(0)=\psi(0)=0,
\quad
\phi'(a)\ge 0,
\quad
\psi'(a)\ge 0,
\]
\[
\phi''(a)\le 0,
\quad
\psi''(a)\le 0.
\]

Moreover,
\[
\psi'(a) \ge \phi'(a)
\quad \text{for all } a\in[0,1],
\]
so automation is weakly more effective on tail exposure.

Thus, automation does not increase exposure and exhibits diminishing
marginal effectiveness in both non-tail and tail risk mitigation.
\end{assumption}

\begin{assumption}[Automation cost convexity]
\label{ass:cost}
The implementation cost function
$K:[0,1]\to\mathbb{R}_+$ is twice continuously differentiable and satisfies
\[
K'(a)\ge 0,
\qquad
K''(a)>0
\quad \text{for all } a\in(0,1).
\]

Furthermore, for each fixed $r$,

\[
K''(a)
>
-\phi''(a)\,B_1(r)
-
\psi''(a)\,B_2(r),
\]

ensuring strict convexity of the total cost function $C(a;r)$.
\end{assumption}

\begin{table}[p]
\centering
\scriptsize
\caption{Model notation (common, static, and dynamic components).}
\label{tab:notation_all}

\renewcommand{\arraystretch}{1.05}
\setlength{\tabcolsep}{3pt}

\begin{adjustbox}{max width=\textwidth,max totalheight=\textheight,keepaspectratio}
\begin{tabular}{p{2.3cm}p{2.2cm}p{8.6cm}}

\hline
Symbol & Type & Description \\
\hline

\multicolumn{3}{l}{Panel A: Common structure} \\
\hline

$\mathbb{N}_0$ & Set & Nonnegative integers (full support). \\
$\mathcal{D}=\{0,\dots,M\}$ & Set & Truncated support for finite-dimensional representation. \\
$M$ & Scalar & Truncation level ($|\mathcal{D}|=M+1$). \\
$\Delta^{M+1}$ & Set & Probability simplex $\{q\in\mathbb{R}_+^{M+1}:\mathbf{1}^\top q=1\}$. \\
$\mathbf{1}$ & Vector & All-ones vector. \\
$\boldsymbol{\ell}$ & Vector & Loss vector; nonnegative and nondecreasing. \\
$d^*$ & Scalar & Tail threshold defining stress region. \\
$\mathsf{I}$ & Operator & Identity operator. \\
$\mathsf{T}$ & Operator & Tail-selection operator $\mathrm{diag}(\mathbf{1}\{d\ge d^*\})$. \\

\hline
\multicolumn{3}{l}{Panel B: Static model (section 3)} \\
\hline

$r>0$ & Scalar & Dispersion parameter (smaller $r$ implies stronger overdispersion). \\
$\mu$ & Scalar & Fixed mean of the distribution. \\
$\theta(r)$ & Scalar & Mean-preserving NB parameterization $\frac{r}{r+\mu}$. \\
$D_r$ & Random variable & NB random variable. \\
$p_d(r)$ & Scalar & NB probability mass function. \\
$Z(r)$ & Scalar & Normalization constant: $Z(r)=\sum_{k=0}^M p_k(r)$. \\
$\pi_d(r)$ & Scalar & Truncated probability: $p_d(r)/Z(r)$. \\
$\boldsymbol{\pi}(r)$ & Vector & Truncated probability vector in $\Delta^{M+1}$. \\
$B_1(r)$ & Scalar & Non-tail exposure. \\
$B_2(r)$ & Scalar & Truncated tail exposure. \\
$a$ & Control & Automation intensity. \\
$\phi(a)$ & Function & Non-tail mitigation (increasing, concave). \\
$\psi(a)$ & Function & Tail mitigation (increasing, concave). \\
$K(a)$ & Function & Automation cost (increasing, convex). \\
$C(a;r)$ & Function & Total static cost. \\
$C_a, C_{aa}$ & Derivatives & First and second derivatives of $C(a;r)$. \\
$F(a,r)$ & Function & First-order condition map $F(a,r)=C_a(a;r)$. \\
$a^*(r)$ & Policy & Static optimal automation. \\
$r^*$ & Scalar & Dispersion threshold solving $C_a(0;r^*)=0$. \\

\hline
\multicolumn{3}{l}{Panel C: Dynamic model (section 4)} \\
\hline

$\mathcal{S}=\{1,\dots,S\}$ & Set & Finite ordered regime space. \\
$\mathcal{X}=\mathcal{S}\times[0,1]$ & Set & State space. \\
$s_t$ & State & Regime at time $t$. \\
$a_t$ & Control & Automation choice at time $t$. \\
$a^{-}$ & State & Previous automation level. \\
$x_t=(s_t,a_{t-1})$ & State & Dynamic state vector. \\
$r_s$ & Scalar & Baseline regime-dependent dispersion. \\
$r_s(a)$ & Function & Automation-dependent dispersion. \\
$\boldsymbol{\pi}(r_s(a))$ & Vector & Truncated NB probability vector. \\
$B_1(s,a)$ & Scalar & Non-tail exposure. \\
$B_2(s,a)$ & Scalar & Truncated tail exposure. \\
$P(a)$ & Kernel & Transition matrix; row-stochastic. \\
$c(s,a)$ & Function & Flow cost. \\
$\Gamma(a,a^{-})$ & Function & Adjustment cost. \\
$\tilde c(s,a,a^{-})$ & Function & Total one-period cost. \\
$\beta\in(0,1)$ & Parameter & Discount factor. \\
$V(s,a^{-})$ & Function & Value function. \\
$\mathcal{T}$ & Operator & Bellman operator. \\
$H(s,a)$ & Function & Smooth Bellman component. \\
$\partial_a H(s,a)$ & Scalar & Marginal derivative of $H$ with respect to $a$. \\
$a^*(s,a^{-})$ & Policy & Dynamic optimal automation. \\
$\underline a(s),\overline a(s)$ & Scalars & Hysteresis band bounds. \\

\hline
\end{tabular}
\end{adjustbox}

\vspace{0.15cm}
\footnotesize
\parbox{0.98\textwidth}{\textit{Note:} All exposure quantities $(B_1,B_2)$ correspond to truncated finite-dimensional approximations of the exact NB model used in the theoretical analysis.}

\end{table}
\newpage
\begin{assumption}[Threshold structure]
\label{ass:threshold}
The derivative of the total cost at zero automation,
\[
C_a(0;r)
=
- \phi'(0)\,B_1(r)
-
\psi'(0)\,B_2(r)
+
K'(0),
\]

is continuous and strictly increasing in $r$.

Moreover, there exist $r_L<r_H$ such that

\[
C_a(0;r_L)<0,
\qquad
C_a(0;r_H)\ge 0.
\]

Then there exists a unique threshold $r^*>0$,
satisfying

\[
a^*(r)=0
\quad \text{for } r\ge r^*,
\qquad
a^*(r)>0
\quad \text{for } r<r^*.
\]
\end{assumption}

For clarity, Table~\ref{tab:notation_all} summarizes the unified notation
for both the static clustered-risk automation model and its dynamic
regime-switching extension. The structure separates
(i) probabilistic primitives governing dispersion,
(ii) linear operator representations of exposure, and
(iii) the automation control problem in both static and dynamic settings.
In particular, dispersion affects the optimization problem only through
the induced probability vector $\boldsymbol{\pi}(r)$ in the static model
and its regime-dependent counterpart $\boldsymbol{\pi}(r_s(a))$ in the dynamic model.
All exposure quantities $(B_1,B_2)$ are defined as linear functionals of these
probability vectors, while persistence and path dependence in the dynamic
model arise through the interaction of the transition kernel and adjustment costs.

\begin{table}[H]
\centering
\caption{Static structural dependency structure.}
\label{tab:static_dependency}
\renewcommand{\arraystretch}{1.2}
\setlength{\tabcolsep}{6pt}
\begin{tabular}{p{4.2cm}p{8.8cm}}
\hline
Object & Structural dependence \\
\hline

NB parameter $\theta(r)$ 
& Mean-preserving mapping $\theta(r)=\frac{r}{r+\mu}$; 
dispersion governed solely by $r$ (mean fixed). \\

Probability vector $\boldsymbol{\pi}(r)$ 
& Truncated and renormalized NB probability vector 
$\boldsymbol{\pi}(r)\in\Delta^{M+1}$ induced by $(r,\mu)$. 
All dispersion effects enter through $\boldsymbol{\pi}(r)$. \\

Non-tail exposure $B_1(r)$ 
& Linear functional $B_1(r)=\boldsymbol{\ell}^\top(\mathsf{I}-\mathsf{T})\boldsymbol{\pi}(r)$. \\

Tail exposure $B_2(r)$ 
& Linear functional $B_2(r)=\boldsymbol{\ell}^\top\mathsf{T}\boldsymbol{\pi}(r)$; 
monotone in dispersion by Lemma~\ref{lem:NB_tail_monotonicity}. \\

Expected loss $L(a;r)$ 
& Affine transformation of $(B_1(r),B_2(r))$:
$L(a;r)=(1-\phi(a))B_1(r)+(1-\psi(a))B_2(r)$. \\

Total cost $C(a;r)$ 
& Composition of exposure mapping and convex automation cost:
$C(a;r)=L(a;r)+K(a)$. \\

First-order condition 
& $C_a(a;r)=-\phi'(a)B_1(r)-\psi'(a)B_2(r)+K'(a)=0$. \\

Optimality map $F(a,r)$ 
& $F(a,r)=C_a(a;r)$; 
dispersion affects $F$ only through $(B_1(r),B_2(r))$. \\

Optimal policy $a^*(r)$ 
& Unique solution of $F(a,r)=0$; 
dispersion influences $a^*(r)$ exclusively via exposure terms. \\

Dispersion threshold $r^*$ 
& Unique value solving $C_a(0;r)=0$, 
separating $a^*(r)=0$ from $a^*(r)>0$. \\

\hline
\end{tabular}
\end{table}

Table~\ref{tab:static_dependency} formalizes the structural flow
of dependence in the static model.
The dispersion parameter $r$ determines the NB parameter
$\theta(r)$ and thereby the probability vector $\pi(r)$.
All exposure components are linear functionals of $\pi(r)$,
while automation enters exclusively through nonlinear mitigation
and convex cost mappings.
Consequently, the comparative statics with respect to dispersion
operate through the exposure pair $(B_1(r),B_2(r))$.

\subsection{Static optimality conditions and structural results}
\label{subsec:static_results}

We now characterize the solution of the static automation problem
\eqref{eq:static_problem} and establish its structural dependence
on dispersion parameter $r$.

By Lemma~\ref{lem:NB_tail_monotonicity}, the exact tail exposure is strictly
decreasing in $r$ under the mean-preserving NB parameterization.
The truncated representation used here preserves this monotonicity for
sufficiently large truncation level.
Hence, stronger overdispersion (smaller $r$) increases expected tail losses
while preserving the mean.

Recall that the total expected cost is
\begin{equation}
C(a;r)
=
(1-\phi(a))\,B_1(r)
+
(1-\psi(a))\,B_2(r)
+
K(a).
\label{eq:C_static_full}
\end{equation}

Differentiating with respect to $a$ yields
\begin{equation}
C_a(a;r)
=
-\phi'(a)\,B_1(r)
-\psi'(a)\,B_2(r)
+K'(a).
\label{eq:FOC_static}
\end{equation}

An interior optimizer $a^*(r)\in(0,1)$ satisfies
\begin{equation}
K'(a^*)
=
\phi'(a^*)\,B_1(r)
+
\psi'(a^*)\,B_2(r).
\label{eq:FOC_balance_static}
\end{equation}

All dependence on dispersion enters exclusively through
$B_1(r)$ and $B_2(r)$.

\begin{lemma}[Strict convexity]
\label{lem:strict_convexity_static}
Under Assumptions~\ref{ass:automation_effectiveness}
and~\ref{ass:cost}, and provided that
\[
K''(a)
>
-\phi''(a)\,B_1(r)
-
\psi''(a)\,B_2(r)
\quad \text{for all } a\in(0,1),\ r>0,
\]
the cost function $C(a;r)$ is strictly convex in $a$
for each fixed $r$.
\end{lemma}

\begin{proof}
The second derivative is
\[
C_{aa}(a;r)
=
- \phi''(a)\,B_1(r)
-
\psi''(a)\,B_2(r)
+
K''(a).
\]
Under the stated curvature condition, $C_{aa}(a;r)>0$ for all $a\in(0,1)$,
which establishes strict convexity.
\end{proof}

\begin{theorem}[Existence and uniqueness]
\label{thm:existence_static}
For each $r>0$, there exists a unique global minimizer
$a^*(r)\in[0,1]$, solving \eqref{eq:static_problem}.
\end{theorem}

\begin{proof}
By Lemma~\ref{lem:strict_convexity_static},
$C(\cdot;r)$ is strictly convex and continuous.
Compactness of $[0,1]$ guarantees existence,
and strict convexity guarantees uniqueness.
\end{proof}

\begin{theorem}[Dispersion monotonicity]
\label{thm:dispersion_monotonicity_static}
Suppose Assumptions~\ref{ass:dispersion}--\ref{ass:cost} hold,
and assume additionally that $B_1'(r)\le 0$ and $B_2'(r)<0$.
On any region where $a^*(r)\in(0,1)$,
\[
\frac{\partial a^*(r)}{\partial r}
=
-\frac{F_r}{F_a},
\quad
F(a,r)=C_a(a;r),
\]
and therefore
\[
\frac{\partial a^*(r)}{\partial r}<0
\quad \text{whenever} \quad
-\psi'(a^*)B_2'(r)
>
\phi'(a^*)|B_1'(r)|.
\]
\end{theorem}

\begin{proof}
At an interior optimum,
$F(a^*(r),r)=0$
and $F_a=C_{aa}(a^*(r);r)>0$.
Differentiating with respect to $r$ gives
\[
F_r
=
-\phi'(a)B_1'(r)
-\psi'(a)B_2'(r).
\]
Under the stated inequality and monotonicity conditions,
$F_r>0$.
The implicit function theorem yields
\[
\frac{\partial a^*(r)}{\partial r}
=
-\frac{F_r}{F_a}
<0.
\]
\end{proof}

\begin{proposition}[Closed-form characterization under quadratic cost]
\label{prop:closed_form_static}
Suppose the control variable affects only the tail component,
so that $\phi(a)\equiv 0$, and let
\[
\psi(a)=a,
\qquad
K(a)=k_1 a^2,
\quad k_1>0.
\]
This specification is consistent with
 Assumptions~\ref{ass:automation_effectiveness}
and~\ref{ass:cost}.
Then the unique minimizer satisfies
\[
a^*(r)
=
\min\left\{1,\frac{B_2(r)}{2k_1}\right\}.
\]
Moreover, there exists a unique threshold $r^*$, satisfying
\[
B_2(r^*)=2k_1,
\]
such that
\[
a^*(r)=
\begin{cases}
1, & r<r^*,\\[6pt]
\dfrac{B_2(r)}{2k_1}, & r\ge r^*.
\end{cases}
\]
The threshold $r^*$ is strictly decreasing in $k_1$.
\end{proposition}

\begin{proof}
Under the stated specialization,
the first-order condition \eqref{eq:FOC_balance_static} reduces to
\[
2k_1 a^* = B_2(r).
\]
Interior solutions therefore satisfy
\(
a^*(r)=B_2(r)/(2k_1).
\)
Imposing the constraint $a\in[0,1]$ yields the stated truncation.
Strict monotonicity of $B_2(r)$ implies the existence and uniqueness
of $r^*$, solving $B_2(r^*)=2k_1$.
Implicit differentiation gives
\[
\frac{dr^*}{dk_1}
=
\frac{2}{B_2'(r^*)}
<0,
\]
since $B_2'(r^*)<0$.
\end{proof}
\begin{theorem}[Dispersion threshold]
\label{thm:threshold_static}
Suppose $B_1(r)$ is nonincreasing and $B_2(r)$ is strictly decreasing,
so that the mapping $r\mapsto C_a(0;r)$ is continuous and strictly increasing.
If there exist $r_L<r_H$ such that
\[
C_a(0;r_L)<0
\quad \text{and} \quad
C_a(0;r_H)\ge 0.
\]
then there exists a unique threshold $r^*>0$ such that
\[
a^*(r)=0 \quad \text{for } r\ge r^*
\quad \text{and} \quad
a^*(r)>0 \quad \text{for } r<r^*.
\]
\end{theorem}

\begin{proof}
Continuity and strict monotonicity imply existence and uniqueness
of $r^*$, satisfying $C_a(0;r^*)=0$.
Strict convexity implies that $a=0$ is optimal if and only if
$C_a(0;r)\ge0$.
\end{proof}

\begin{figure}[htbp]
\centering
\includegraphics[width=1.0\linewidth]{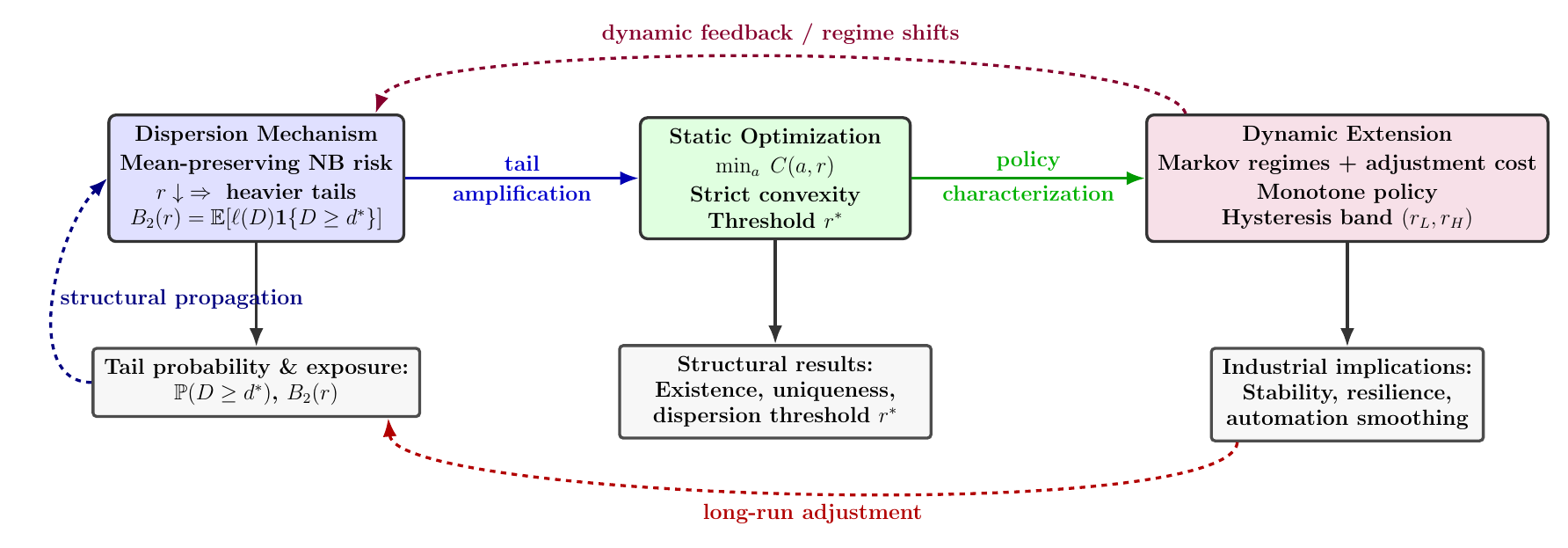}
\caption{
Structural architecture of the dispersion--threshold mechanism.
Solid arrows represent core optimization logic,
while dashed arrows indicate interpretative dynamic feedback.
}
\label{fig:structure}
\end{figure}

To visualize the structural transition from dispersion-driven static optimization
to the dynamic regime-switching extension, Figure~\ref{fig:structure}
summarizes the logical architecture of the model.

\section{Dynamic regime-switching model with adjustment frictions}
\label{sec:dynamic}

In Section~\ref{sec:static}, we established that dispersion-sensitive tail exposure
induces a threshold structure in optimal automation.
We now embed the static environment into a dynamic regime-switching model
with costly adjustment, generating endogenous hysteresis.

Automation becomes a persistent state variable,
and clustered-risk shocks propagate intertemporally
through regime persistence and adjustment frictions.

\subsection{State space, regime-dependent dispersion, and structural feedback}
\label{subsec:dynamic_state}

Time is discrete, $t=0,1,2,\dots$.
In contrast to the static environment of Section~\ref{sec:static},
automation now becomes a persistent state variable.
At each date $t$, the economy is characterized by the state vector
\[
x_t=(s_t,a_{t-1}),
\]
where $s_t\in\mathcal{S}$ denotes the current risk regime and
$a_{t-1}\in[0,1]$ is previously installed automation intensity.

Let $\mathcal{S}=\{1,2,\dots,S\}$ be a finite ordered set equipped with
its natural order, where larger indices correspond to more severe regimes.


In the static model, clustered risk is summarized by a scalar dispersion
parameter $r>0$, with tail exposure decreasing in $r$ under mean preservation.
The dynamic extension replaces this scalar by a regime-dependent family
$\{r_s\}_{s\in\mathcal{S}}$.

Each regime $s$ is associated with $r_s>0$,  satisfying
\begin{equation}
r_1 > r_2 > \dots > r_S.
\label{eq:regime_dispersion_order}
\end{equation}

Since the mean $\mu$ is fixed,
smaller $r_s$ implies larger variance
\[
\mathrm{Var}(D)=\mu+\frac{\mu^2}{r_s},
\]
and therefore stronger overdispersion and heavier tail clustering.

Each regime induces a truncated probability vector
\[
\boldsymbol{\pi}(r_s)\in\Delta^{M+1}.
\]

Define exposure components
\begin{equation}
B_1(r_s)
=
\boldsymbol{\ell}^\top(\mathsf{I}-\mathsf{T})\,\boldsymbol{\pi}(r_s),
\qquad
B_2(r_s)
=
\boldsymbol{\ell}^\top\mathsf{T}\,\boldsymbol{\pi}(r_s).
\label{eq:dynamic_exposure_def}
\end{equation}

By Lemma~\ref{lem:NB_tail_monotonicity}, the exact tail exposure is strictly decreasing in $r$.
The truncated representation used here preserves this ordering for sufficiently large $M$.
Hence,
\begin{equation}
B_2(r_1) \le B_2(r_2) \le \dots \le B_2(r_S).
\label{eq:regime_tail_order}
\end{equation}

Thus, higher-index regimes exhibit strictly larger tail exposure.


Automation modifies dispersion through $r_s(a)$ with
\[
r_s'(a)\ge 0.
\]

Exposure becomes
\[
B_j(s,a)=B_j(r_s(a)), \quad j=1,2.
\]

Since the exact tail exposure is decreasing in $r$,
and the truncated representation preserves this property,
the chain rule yields
\begin{equation}
\frac{\partial}{\partial a} B_2(r_s(a))
=
\frac{\partial B_2}{\partial r}(r_s(a))\, r_s'(a)
\le 0.
\label{eq:chainrule_tail}
\end{equation}

Thus automation weakly reduces tail exposure within each regime.


Automation also affects regime transitions via
\[
P(a)=\bigl[P_{ss'}(a)\bigr],
\qquad
\sum_{s'}P_{ss'}(a)=1.
\]

\begin{assumption}[Automation structure]
\label{ass:dynamic_automation}
For each $s\in\mathcal{S}$, $r_s(a)$ is continuously differentiable,
bounded, and satisfies $r_s'(a)\ge 0$.

For each $s,s'$, $P_{ss'}(a)$ is continuously differentiable in $a$.

Moreover, for each fixed $s$ and all $a'\ge a$,
the transition distributions satisfy first-order stochastic dominance:
\[
\sum_{s'\ge k} P_{ss'}(a')
\;\ge\;
\sum_{s'\ge k} P_{ss'}(a)
\quad \text{for all } k.
\]
\end{assumption}

\subsection{Per-period cost and adjustment frictions}
\label{subsec:dynamic_cost}

Given the current regime $s\in\mathcal{S}$ and previously installed automation level
$a^{-}\in[0,1]$, the firm chooses current automation $a\in[0,1]$.
Automation is therefore an endogenous persistent state variable, and the state space is
\[
\mathcal{X}=\mathcal{S}\times[0,1],
\qquad
x=(s,a^{-}).
\]

The functions $\phi,\psi:[0,1]\to[0,1]$ represent automation effectiveness in reducing
non-tail and tail exposure, respectively.

\begin{assumption}[Mitigation functions]
\label{ass:dynamic_mitigation}
The functions $\phi$ and $\psi$ are twice continuously differentiable on $[0,1]$ and satisfy
\[
\phi(0)=\psi(0)=0,
\qquad
\phi'(a)\ge 0,
\qquad
\psi'(a)\ge 0,
\]
\[
\phi''(a)\le 0,
\qquad
\psi''(a)\le 0.
\]
Moreover, $\psi'(a)\ge \phi'(a)$ for all $a\in[0,1]$.
\end{assumption}

Conditional on regime $s$, the contemporaneous operating cost is
\begin{equation}
c(s,a)
=
B_1(s,a)\bigl(1-\phi(a)\bigr)
+
B_2(s,a)\bigl(1-\psi(a)\bigr)
+
K(a),
\label{eq:dynamic_flow_cost}
\end{equation}
where $K:[0,1]\to\mathbb{R}_+$ is the implementation cost, and we use the shorthand
$B_j(s,a):=B_j\!\bigl(r_s(a)\bigr)$ for $j\in\{1,2\}$.

\begin{assumption}[Convexity and increasing differences]
\label{ass:dynamic_convex_id}
The function $c(s,a)$ is twice continuously differentiable in $a$.
For each $s\in\mathcal{S}$, the mapping $a\mapsto c(s,a)$ is strictly convex on $[0,1]$.
Moreover, $c(s,a)$ has increasing differences in $(s,a)$ with respect to the natural order on
$\mathcal{S}$ and the usual order on $[0,1]$.
\end{assumption}

Automation changes incur costs
\begin{equation}
\Gamma(a,a^{-})
=
\kappa_1 |a-a^{-}|
+
\frac{\kappa_2}{2}(a-a^{-})^2,
\label{eq:adjustment_cost}
\end{equation}
with $\kappa_1,\kappa_2\ge 0$.
The function $\Gamma$ is continuous on $[0,1]^2$ and convex in $a$ for each fixed $a^{-}$.

Define
\begin{equation}
\tilde{c}(s,a,a^{-})
=
c(s,a)
+
\Gamma(a,a^{-}).
\label{eq:total_flow_cost}
\end{equation}

\begin{assumption}[Bounded cost]
\label{ass:bounded_cost}
The function $\tilde{c}(s,a,a^{-})$ is bounded on $\mathcal{S}\times[0,1]^2$.
\end{assumption}

For each fixed $(s,a^{-})$, the mapping $a\mapsto \tilde{c}(s,a,a^{-})$ is continuous
and strictly convex on $[0,1]$.

Let $\beta\in(0,1)$.
Given state $(s,a^{-})$, choosing $a$ induces the next-period regime distribution via $P(a)$
and sets the next automation state equal to $a$.
The value function satisfies
\begin{equation}
V(s,a^{-})
=
\min_{a\in[0,1]}
\left\{
\tilde{c}(s,a,a^{-})
+
\beta
\sum_{s'\in\mathcal{S}} P_{ss'}(a)\,V(s',a)
\right\}.
\label{eq:bellman}
\end{equation}

Define the Bellman operator
\[
(\mathcal{T}V)(s,a^{-})
=
\min_{a\in[0,1]}
\left\{
\tilde{c}(s,a,a^{-})
+
\beta
\sum_{s'\in\mathcal{S}} P_{ss'}(a)\,V(s',a)
\right\}.
\]

\begin{proposition}[Bellman contraction and fixed-point uniqueness]
\label{prop:dynamic_contraction}
Under Assumption~\ref{ass:bounded_cost}, row-stochasticity of $P(a)$ for each $a\in[0,1]$,
and $\beta\in(0,1)$, operator $\mathcal{T}$ is a contraction on
$\mathcal{B}(\mathcal{X})$ under the sup norm with modulus $\beta$:
\[
\|\mathcal{T}V-\mathcal{T}W\|_\infty
\le
\beta \|V-W\|_\infty.
\]
Hence, there exists a unique bounded fixed point $V$, satisfying $V=\mathcal{T}V$.
\end{proposition}

\begin{proof}
For any bounded functions $V,W$ and any $(s,a^{-})\in\mathcal{X}$, for each $a\in[0,1]$,
\[
\left|
\sum_{s'\in\mathcal{S}} P_{ss'}(a)V(s',a)
-
\sum_{s'\in\mathcal{S}} P_{ss'}(a)W(s',a)
\right|
\le
\|V-W\|_\infty,
\]
since $P(a)$ is row-stochastic. Therefore, for every admissible $a$,
\[
\left|
\tilde{c}(s,a,a^{-})
+
\beta \sum_{s'\in\mathcal{S}} P_{ss'}(a)V(s',a)
-
\tilde{c}(s,a,a^{-})
-
\beta \sum_{s'\in\mathcal{S}} P_{ss'}(a)W(s',a)
\right|
\le
\beta\|V-W\|_\infty.
\]
Taking the minimum over $a\in[0,1]$ preserves nonexpansiveness, yielding
\[
\|\mathcal{T}V-\mathcal{T}W\|_\infty
\le
\beta\|V-W\|_\infty.
\]
Hence, $\mathcal{T}$ is a contraction. Since $\beta\in(0,1)$, and
$\mathcal{B}(\mathcal{X})$ is complete under the sup norm, the Banach fixed-point theorem
implies existence and uniqueness of a bounded fixed point $V$, satisfying $V=\mathcal{T}V$.
\end{proof}

\subsection{Dynamic structural results}
\label{subsec:dynamic_results}

We now establish the structural properties of the dynamic automation problem.
Throughout this subsection, let the state space be
$\mathcal{X}=\mathcal{S}\times[0,1]$, where $\mathcal{S}$
is a finite totally ordered set of regimes, and $a^{-}\in[0,1]$
denotes previously installed automation.

\begin{assumption}[Dynamic primitives]
\label{ass:dynamic_primitives}
The discount factor satisfies $\beta\in(0,1)$.
The per-period cost $c(s,a)$ is twice continuously differentiable
and strictly convex in $a$ for each $s\in\mathcal{S}$.
The adjustment cost $\Gamma(a,a^{-})$ is continuous on $[0,1]^2$
and convex in both arguments.
The transition kernel $P(a)=[P_{ss'}(a)]$ is row-stochastic
and continuous in $a$.
\end{assumption}

\begin{assumption}[Increasing differences]
\label{ass:dynamic_ID}
The flow cost $c(s,a)$ has increasing differences in $(s,a)$.
Moreover, for any bounded function $W(s,a^{-})$
that is nondecreasing in $s$ for each fixed $a^{-}$,
the continuation mapping
\[
(s,a)\mapsto
\sum_{s'\in\mathcal{S}} P_{ss'}(a)\, W(s',a)
\]
has increasing differences in $(s,a)$.
\end{assumption}

Assumption~\ref{ass:dynamic_ID} ensures that the transition kernel
preserves regime monotonicity in a manner compatible with
monotone comparative statics.

\begin{lemma}[Existence, uniqueness, and regularity]
\label{lem:dynamic_existence}
The Bellman operator
\[
(\mathcal{T}V)(s,a^{-})
=
\min_{a\in[0,1]}
\Bigl\{
c(s,a)
+
\Gamma(a,a^{-})
+
\beta \sum_{s'\in\mathcal{S}} P_{ss'}(a)\,V(s',a)
\Bigr\}
\]
is a contraction on $\mathcal{B}(\mathcal{X})$
under the sup norm with modulus $\beta$.
Hence, there exists a unique bounded continuous value function
$V$, satisfying $V=\mathcal{T}V$.
Furthermore, $V(s,a^{-})$ is nondecreasing in $s$.
\end{lemma}

\begin{proof}
For any bounded $V_1,V_2$,
\[
\left|
\sum_{s'} P_{ss'}(a)V_1(s',a)
-
\sum_{s'} P_{ss'}(a)V_2(s',a)
\right|
\le
\|V_1-V_2\|_\infty,
\]
since $P(a)$ is row-stochastic.
Taking the minimum over $a\in[0,1]$
preserves nonexpansiveness,
yielding
\[
\|\mathcal{T}V_1-\mathcal{T}V_2\|_\infty
\le
\beta \|V_1-V_2\|_\infty.
\]
Therefore, $\mathcal{T}$ is a contraction.

To show monotonicity in $s$,
initialize $V^{(0)}(s,a^{-})=0$.
Suppose $V^{(n)}$ is nondecreasing in $s$.
By Assumption~\ref{ass:dynamic_ID},
the continuation term is nondecreasing in $s$.
The objective inside the minimization
is therefore nondecreasing in $s$.
Minimization over $a$ preserves monotonicity.
Thus, $V^{(n+1)}$ is nondecreasing in $s$.
Convergence of value iteration implies
$V$ is nondecreasing in $s$.
\end{proof}

\begin{theorem}[Monotone automation in regime]
\label{thm:monotone_policy}
Under Assumptions~\ref{ass:dynamic_primitives}
and~\ref{ass:dynamic_ID},
the optimal policy satisfies
\[
a^*(s',a^{-}) \ge a^*(s,a^{-})
\quad
\text{whenever }
s'\succeq s.
\]
\end{theorem}

\begin{proof}
Fix $a^{-}$ and define the objective
\[
\Psi(s,a)
=
c(s,a)
+
\Gamma(a,a^{-})
+
\beta \sum_{s'} P_{ss'}(a)V(s',a).
\]
By Assumption~\ref{ass:dynamic_ID},
both the flow component and the continuation component
have increasing differences in $(s,a)$.
Since $\Gamma(a,a^{-})$ does not depend on $s$,
$\Psi(s,a)$ has increasing differences in $(s,a)$.

The feasible set $[0,1]$ is a compact lattice,
and strict convexity in $a$ ensures a unique minimizer.
Topkis' monotonicity theorem
then implies that the argmin correspondence
is increasing in $s$.
\end{proof}

\begin{theorem}[Dynamic hysteresis band]
\label{thm:hysteresis}
Suppose $\Gamma(a,a^{-})=\kappa_1|a-a^{-}|$
with $\kappa_1>0$.
Define
\[
H(s,a)
=
c(s,a)
+
\beta \sum_{s'\in\mathcal{S}} P_{ss'}(a)V(s',a).
\]
If $H(s,a)$ is continuously differentiable
and strictly convex in $a$,
then for each regime $s$
there exist thresholds
$\underline a(s)\le \overline a(s)$
such that
\[
a^*(s,a^{-})
=
\begin{cases}
\underline a(s), & a^{-}<\underline a(s),\\
a^{-}, & a^{-}\in[\underline a(s),\overline a(s)],\\
\overline a(s), & a^{-}>\overline a(s).
\end{cases}
\]
The inaction region satisfies
\[
\big|\partial_a H(s,a^{-})\big|
\le
\kappa_1.
\]
\end{theorem}

\begin{proof}
Fix $(s,a^{-})$ and define
\[
J(a)=H(s,a)+\kappa_1|a-a^{-}|.
\]
Since $H$ is strictly convex and $|a-a^{-}|$ is convex,
$J$ is strictly convex and has a unique minimizer.

For $a>a^{-}$,
\[
\partial_a J(a)
=
\partial_a H(s,a)+\kappa_1,
\]
and for $a<a^{-}$,
\[
\partial_a J(a)
=
\partial_a H(s,a)-\kappa_1.
\]
At $a=a^{-}$,
\[
\partial J(a^{-})
=
[\partial_a H(s,a^{-})-\kappa_1,\,
 \partial_a H(s,a^{-})+\kappa_1].
\]
Thus, $a=a^{-}$ is optimal iff
\[
|\partial_a H(s,a^{-})|\le\kappa_1.
\]

Since $H(s,a)$ is strictly convex in $a$,
its derivative $\partial_a H(s,a)$ is strictly increasing.
This implies the existence of unique
$\underline a(s)$ and $\overline a(s)$.
\end{proof}

\begin{theorem}[Crisis persistence and path dependence]
\label{thm:crisis_persistence}
Suppose Assumptions~\ref{ass:dynamic_primitives}
and~\ref{ass:dynamic_ID} hold.
Assume linear adjustment cost
\[
\Gamma(a,a^{-})=\kappa_1 |a-a^{-}|,
\qquad \kappa_1>0,
\]
and let
\[
H(s,a)
=
c(s,a)
+
\beta \sum_{s'\in\mathcal{S}} P_{ss'}(a)V(s',a)
\]
be continuously differentiable and strictly convex in $a$
for each regime $s$.

Let $s_L \prec s_H$ denote a calm and a stress regime.
Let $a_H = a^*(s_H,a^{-})$ denote the automation level chosen in regime $s_H$
for some previous installation $a^{-}$.
Let $a_L^{stat}$ denote the static optimizer in regime $s_L$,
defined as the unique solution of
\[
\partial_a c(s_L,a_L^{stat})=0.
\]

Suppose
\[
a_H > a_L^{stat}
\]
and the calm-regime dynamic marginal incentive to reduce automation
at $a_H$ satisfies
\begin{equation}
\partial_a H(s_L,a_H) < \kappa_1.
\label{eq:persistence_condition}
\end{equation}

Then the post-transition optimal policy satisfies
\[
a^*(s_L,a_H) \ge a_H > a_L^{stat}.
\]

In particular, the upward automation adjustment taken in the stress regime
persists strictly above the calm-regime static optimum
for at least one period after the transition.
\end{theorem}

\begin{proof}
Fix the post-transition state $(s_L,a^{-})=(s_L,a_H)$.
Define the objective
\[
J(a)
=
H(s_L,a)
+
\kappa_1 |a-a_H|.
\]

Since $H(s_L,a)$ is strictly convex
and $|a-a_H|$ is convex,
$J$ is strictly convex and admits a unique minimizer,
denoted $a_L^*:=a^*(s_L,a_H)$.

We show that $a_L^*\ge a_H$.

Suppose instead that $a_L^*<a_H$.
On the region $a<a_H$,
the objective is differentiable with
\[
\partial_a J(a)
=
\partial_a H(s_L,a)
-
\kappa_1.
\]
The first-order condition for an interior minimizer in this region requires
\[
\partial_a H(s_L,a_L^*)=\kappa_1.
\]

Since $H(s_L,\cdot)$ is strictly convex,
$\partial_a H(s_L,a)$ is strictly increasing in $a$.
Because $a_L^*<a_H$,
we have
\[
\partial_a H(s_L,a_L^*)
<
\partial_a H(s_L,a_H).
\]
Thus,
\[
\kappa_1
=
\partial_a H(s_L,a_L^*)
<
\partial_a H(s_L,a_H),
\]
which contradicts condition~\eqref{eq:persistence_condition}.

Therefore, $a_L^*\ge a_H$.
Finally, since $a_H>a_L^{stat}$ by hypothesis,
it follows that
\[
a^*(s_L,a_H)
=
a_L^*
\ge
a_H
>
a_L^{stat}.
\]
\end{proof}

The unified notation table introduced in Section~\ref{sec:static}
(Table~\ref{tab:notation_all}, Panel C) summarizes the core objects of the dynamic
regime-switching model. The structure separates (i) regime-dependent dispersion
primitives $r_s(a)$ and the induced probability vector $\boldsymbol{\pi}(r_s(a))$,
(ii) linear exposure operators $(B_1,B_2)$ that transmit dispersion into losses,
(iii) intertemporal propagation through the automation-dependent transition
kernel $P(a)$, and (iv) adjustment frictions via $\Gamma(a,a^{-})$. This dynamic structure builds directly on the static formulation in Section~\ref{sec:static},
extending the dispersion-driven exposure mapping to a regime-dependent and intertemporal setting.

\begin{table}[H]
\centering
\caption{Dynamic structural dependency structure.}
\label{tab:dynamic_dependency}
\renewcommand{\arraystretch}{1.28}
\setlength{\tabcolsep}{6pt}
\begin{tabular}{p{4.4cm}p{7.3cm}}
\hline
Object & Structural dependence \\
\hline

Regime evolution $s_{t+1}$ 
& Controlled markov transition:
$\mathbb{P}(s_{t+1}=s'|s_t=s,a_t=a)=P_{ss'}(a)$. \\

Within-regime dispersion 
& Automation shifts dispersion via $r_s(a)$,
altering $\boldsymbol{\pi}(r_s(a))$ and tail clustering. \\

Probability vector $\boldsymbol{\pi}(r_s(a))$ 
& Determined by $r_s(a)$ (mean fixed); 
all dispersion effects enter through this vector. \\

Exposure $(B_1,B_2)$ 
& Linear operators applied to $\boldsymbol{\pi}(r_s(a))$. \\

Flow cost $c(s,a)$ 
& Depends on exposure and mitigation functions
$(B_1,B_2,\phi,\psi,K)$. \\

Adjustment cost $\Gamma(a,a^{-})$ 
& Depends on $(a,a^{-})$ and friction parameters $(\kappa_1,\kappa_2)$. \\

Bellman operator $\mathcal{T}$ 
& Defined by $(c,\Gamma,\beta,P(\cdot))$. \\

Value function $V$ 
& Unique contraction fixed point of $\mathcal{T}$. \\

Dynamic marginal $\partial_a H(s,a)$ 
& Includes:
static marginal cost,
continuation-value derivative,
and transition-feedback term. \\

Policy $a^*(s,a^{-})$ 
& Determined jointly by dispersion mitigation $r_s(a)$
and transition kernel $P(a)$. \\

Hysteresis band 
& Characterized by $|\partial_a H(s,a^{-})|\le \kappa_1$. \\

Persistence effect 
& Arises from regime persistence in $P(a)$
combined with adjustment frictions. \\

\hline
\end{tabular}
\end{table}
Table~\ref{tab:dynamic_dependency} formalizes the structural flow of dependence
in the dynamic model. Within each regime, automation shifts dispersion through
$r_s(a)$, thereby modifying the probability vector
$\boldsymbol{\pi}(r_s(a))$ and the exposure pair $(B_1,B_2)$.
Intertemporally, automation influences the regime process through the
controlled transition kernel $P(a)$. Consequently, dynamic incentives operate through two distinct channels:
A within-regime dispersion-mitigation channel and a regime-transition channel.
Adjustment frictions then generate hysteresis by preventing instantaneous
realignment with the static optimum.

\providecommand{\req}{\checkmark}
\providecommand{\aux}{\(\circ\)}

\begin{table}[!htpb]
\centering
\scriptsize
\begin{adjustbox}{angle=90}
\begin{minipage}{\textheight}
\caption{
Structural dependency of assumptions across static and dynamic theoretical results.
$\checkmark$ = directly required; $\circ$ = auxiliary.
}
\label{tab:structural-dependency}
\renewcommand{\arraystretch}{1.2}
\setlength{\tabcolsep}{3pt}

\begin{tabular}{lcccccccccc p{4.5cm}}
\hline
Assumption
& \rotatebox{80}{Lemma S1: Tail convexity}
& \rotatebox{80}{Prop S2: Threshold existence}
& \rotatebox{80}{Thm S3: Uniqueness}
& \rotatebox{80}{Cor S4: Comparative statics}
& \rotatebox{80}{Prop D1: Regime value}
& \rotatebox{80}{Thm D2: Policy existence}
& \rotatebox{80}{Thm D3: Monotonicity}
& \rotatebox{80}{Prop D4: Hysteresis}
& \rotatebox{80}{Thm D5: Stability}
& \rotatebox{80}{Cor D6: Regret bound}
& Notes \\
\hline

\multicolumn{11}{l}{Assumptions} \\

A1: NB dispersion structure
  & \req & \req & \aux & \aux & \aux & \aux & \aux & \aux & \aux & \req
  & Determines tail amplification in the exact model, preserved under truncation \\

A2:  Convex order / stochastic ordering
  & \req & \aux & \aux & \req & \aux & \aux & \req & \aux & \aux & \aux
  & Ensures ordering of tail risk and monotone policy response \\

A3: Concave mitigation ($\phi'',\psi'' \le 0$)
  & \aux & \req & \req & \aux & \aux & \aux & \aux & \aux & \aux & \aux
  & Drives strict convexity of the static objective \\

A4: Strict convex automation cost ($K''>0$)
  & \aux & \req & \req & \aux & \aux & \aux & \aux & \aux & \aux & \aux
  & Guarantees uniqueness of optimal automation \\

A5: Threshold derivative condition
  & \aux & \aux & \aux & \req & \aux & \aux & \aux & \aux & \aux & \aux
  & Pins down the dispersion cutoff $r^*$ \\

A6: Regime persistence (Markov structure)
  &  &  &  &  & \req & \req & \aux & \aux & \req & \aux
  & Generates dynamic amplification and persistence effects \\

A7: Adjustment friction
  &  &  &  &  & \aux & \aux & \aux & \req & \aux & \aux
  & Produces hysteresis band and inaction region \\

A8: Increasing differences (supermodularity)
  &  &  &  &  & \aux & \req & \req & \aux & \aux & \aux
  & Ensures monotone dynamic policy (Topkis theorem) \\

A9: Discounting and boundedness ($\beta<1$)
  &  &  &  &  & \req & \req & \aux & \aux & \req & \aux
  & Ensures Bellman contraction and value existence \\

A10: Transition monotonicity
  &  &  &  &  & \aux & \req & \req & \aux & \aux & \aux
  & Supports regime-order monotonicity \\

A11: Compact action set
  & \req & \req & \req & \req & \req & \req & \req & \req & \req & \req
  & Technical regularity ensuring existence of minimizers \\

\hline
\end{tabular}
\end{minipage}
\end{adjustbox}
\end{table}

To clarify the structural role of the maintained assumptions across the
static and dynamic components of the model,
Table~\ref{tab:structural-dependency} maps each assumption to the
lemmas, propositions, and theorems it directly supports.
Moreover, the table distinguishes assumptions that are indispensable for
existence, uniqueness, and dispersion-threshold results in the static model,
and those that become essential only in the dynamic extension for establishing
monotonicity, hysteresis, and stability properties.
This structural mapping makes transparent the minimality and non-redundancy
of the maintained assumptions and clarifies the logical architecture of the theory.

\begin{figure}[htbp]
\centering
\includegraphics[width=0.92\linewidth]{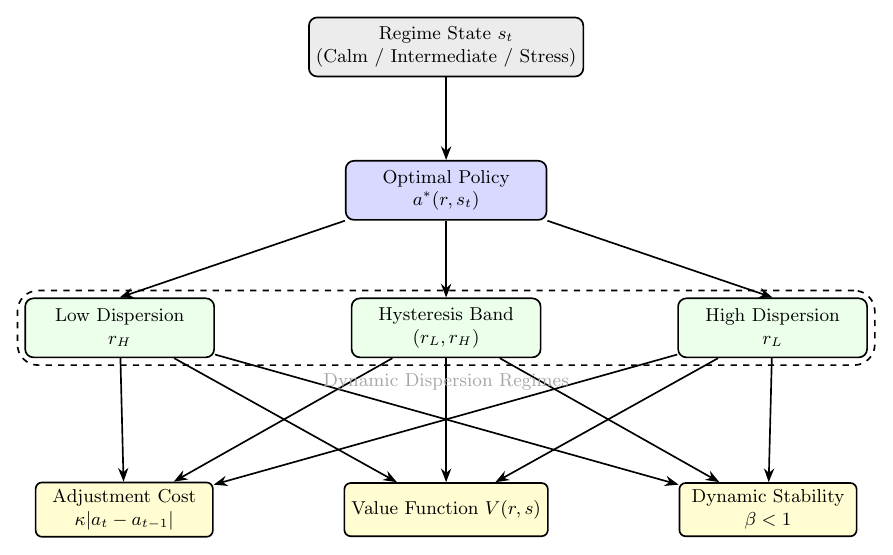}
\caption{
Dynamic policy geometry under regime persistence.
The hysteresis band $(r_L,r_H)$ arises from switching costs,
creating an inaction region in the dispersion space.
}
\label{fig:hysteresis}
\end{figure}

\newpage
Figure~\ref{fig:hysteresis} illustrates the geometry of the optimal policy
under adjustment frictions, highlighting the endogenous hysteresis band.

The numerical exercises that follow are designed solely to illustrate
the structural comparative statics implied by the theory.
They serve to visualize the underlying mechanisms and do not introduce
any assumptions beyond those formally stated above.

\subsection{Numerical solution of the bellman problem}
\label{subsec:dynamic_numerics}

The theoretical results above characterize the structure of the dynamic
automation problem through the Bellman equation in
\eqref{eq:bellman}. To compute the policy used in the numerical
illustrations of Sections~5 and~6, we solve this Bellman problem on a
discretized state space.

The regime space $\mathcal{S}$ is finite and directly enumerated.
The automation state and control spaces are discretized on a uniform grid
\[
\mathcal{A}=\left\{0,\frac{1}{N_a-1},\frac{2}{N_a-1},\dots,1\right\},
\]
with baseline choice $N_a=51$.
Hence, the previously installed automation level $a^{-}$ and the
current decision $a$ are evaluated on the same compact grid.

For each regime $s\in\mathcal{S}$ and action $a\in\mathcal{A}$, the
automation-dependent dispersion parameter $r_s(a)$ determines the
truncated probability vector $\boldsymbol{\pi}(r_s(a))$.
All regime-contingent exposure terms are then computed as
\[
B_1(s,a)=\boldsymbol{\ell}^\top(\mathsf{I}-\mathsf{T})\boldsymbol{\pi}(r_s(a)),
\qquad
B_2(s,a)=\boldsymbol{\ell}^\top\mathsf{T}\boldsymbol{\pi}(r_s(a)).
\]
The NB distribution is truncated at a sufficiently large
finite level $M$ so that truncation error is numerically negligible.
This preserves the theoretical role of the exact model while enabling a
finite-dimensional implementation.

Given these objects, the Bellman operator is evaluated pointwise on the
discretized state space as
\[
(\mathcal{T}V)(s,a^{-})
=
\min_{a\in\mathcal{A}}
\left\{
\tilde{c}(s,a,a^{-})
+
\beta\sum_{s'\in\mathcal{S}}P_{ss'}(a)V(s',a)
\right\}.
\]

We initialize value iteration with $V^{(0)}(s,a^{-})=0$ for all
$(s,a^{-})\in\mathcal{S}\times\mathcal{A}$ and update recursively by
\[
V^{(m+1)}=\mathcal{T}V^{(m)}.
\]
Iteration is terminated when the sup-norm difference satisfies
\[
\|V^{(m+1)}-V^{(m)}\|_{\infty}<\varepsilon,
\]
where $\varepsilon=10^{-6}$ in the baseline implementation.

Because the action set is compact and discretized directly on $[0,1]$,
boundary treatment is handled explicitly by including the endpoints
$a=0$ and $a=1$ in the grid. No extrapolation outside the admissible
control region is required. The computed policy
$a^*(s,a^{-})$ is obtained as the pointwise minimizer of the discretized
Bellman operator. The numerical results in Section~5 are based on this
solution procedure, In Section~5, we report the resulting structural
validation and in Section~6, we provide the empirical illustration.

\section{Synthetic structural experiments}
\label{sec:synthetic_experiments}

All numerical experiments are conducted within a fully controlled
synthetic regime-switching economy constructed to mirror the structural
assumptions of the model. This synthetic environment is not intended for
empirical calibration, but is designed to provide a transparent and
theory-consistent setting in which the structural mechanisms of the
model, such as dispersion effects, tail amplification, and regime-dependent
dynamics, can be systematically isolated and examined. Demand (or shock
counts) follows a mean-preserving NB distribution with
fixed mean $\mu = 4$ and dispersion parameter $r$. Lower values of $r$
imply heavier tails and stronger overdispersion. For numerical stability,
the distribution is truncated at $D \le 80$; the resulting normalization
error remains below $10^{-15}$ across the parameter grid.

The dynamic environment contains three regimes, Calm, Intermediate, and Stress, with baseline dispersion levels
$r_1 > r_2 > r_3$, respectively.
The control variable affects effective dispersion according to
\[
r_s(a) = \min\{ r_{s,\text{base}} + \delta_s a,\, r_{s,\max} \},
\]
preserving the mean while attenuating tail risk.
Regime transitions follow a Markov kernel $P(a)$.
Higher control intensity reallocates probability mass toward safer regimes,
shifting transitions from Stress to Intermediate and from
Intermediate to Calm while preserving stochastic monotonicity.

The parameter values are selected to ensure consistency with the
structural roles defined in Sections~\ref{sec:static} and~\ref{sec:dynamic},
rather than to fit a dataset. In particular, the dispersion grid
spans high- and low-dispersion regimes under the mean-preserving
specification, ensuring that both corner and interior solutions arise in
the static optimization problem. Cost parameter $k_1$ is calibrated so
that the marginal cost of control is comparable to the marginal reduction
in tail exposure, thereby generating a nontrivial dispersion threshold $r^*$.

The discount factor $\beta=0.97$ is chosen to reflect persistent but
discounted continuation effects, so that future regime dynamics affect the
policy without overwhelming current-period costs. The adjustment-cost
parameters $\kappa_1$ and $\kappa_2$ are selected to generate gradual
rather than instantaneous adjustment, enabling the model to exhibit an
economically meaningful inaction region and partial hysteresis. These
parameters jointly determine the sensitivity of the policy to dispersion,
regime persistence, and lagged control. Their effects are therefore not
interpreted in isolation, but through the threshold, monotonicity, and
forward-simulation exercises reported below.

The dynamic program is solved on a discretized action space
$a \in [0,1]$ with 51 grid points. Given the maintained discount factor
$\beta = 0.97$, the model incorporates convex adjustment cost
\[
\kappa_1 |a - a_{prev}| + \frac{1}{2}\kappa_2 (a - a_{prev})^2.
\]
The solution is obtained via value iteration, with convergence evaluated
under the sup-norm criterion
\[
\|V^{(k+1)} - V^{(k)}\|_\infty < 10^{-5}.
\]
Because the control space is compact and includes both boundary points,
boundary conditions are handled explicitly without extrapolation.

The resulting policy is used to simulate a 20-year
business-day panel (approximately 5040 observations),
initialized in the calm regime.

Returns combine diffusion and jump components:
\[
r_t = \mu_s + \sigma_s \varepsilon_t - \lambda_s D_t,
\]
where $\varepsilon_t$ follows a Student-$t$ distribution.

All numerical experiments are implemented in Python. Data processing and 
statistical estimation are conducted using standard scientific libraries, 
including NumPy and pandas, while numerical optimization and dynamic 
programming procedures are implemented using custom scripts. A fixed global 
random seed is used throughout to ensure full reproducibility.

\subsection{Mean-preserving dispersion and tail amplification}
\label{subsec:mpd_tail}

We first illustrate the key structural mechanism established in
Lemma~\ref{lem:NB_tail_monotonicity}.
This experiment corresponds to the static model in
Section~\ref{sec:static}, where dispersion $r$ is treated as an
exogenous parameter, and no intertemporal dynamics are present.
Under mean-preserving dispersion (MPD),
an increase in dispersion (equivalently, a decrease in the
NB shape parameter $r$)
strictly amplifies tail exposure.

Demand $D$ follows a NB distribution with fixed mean 
$\mu = 4$ and dispersion parameter $r$.
The probability mass function is truncated at $k \le 200$
to ensure bounded support.
The loss function is monotone and bounded:
\[
\ell(k) = \min\{k^{1.2},\,200\}.
\]

Tail exposure is defined as
\[
B_2(r)
= \mathbb{E}\big[\ell(D)\mathbf{1}\{D \ge d^*\}\big],
\qquad d^* = 10.
\]

Over the grid $r \in [1.5,40]$,
both tail exposure and tail probability are strictly decreasing in $r$:
\[
\frac{\partial B_2(r)}{\partial r} < 0,
\qquad
\frac{\partial \mathbb{P}(D \ge d^*)}{\partial r} < 0.
\]

The NB distribution is parameterized such that the mean
remains fixed at $\mu=4$ for all values of $r$.
The truncated mean satisfies
\[
\max_r \left| \mathbb{E}_{\text{trunc}}[D] - \mu \right| < 10^{-12},
\]
confirming that the numerical implementation preserves the MPD property.

Truncation error is numerically negligible:
\[
\max_{r} \left|1 - Z_{\text{trunc}}(r)\right|
= 1.332 \times 10^{-15}.
\]
Increasing the truncation bound from $200$ to $400$
changes $B_2(r)$ by less than $10^{-6}$ across all $r$,
confirming that bounded support does not affect tail behavior.

Table~\ref{tab:static_grid_summary} reports representative grid values.
By construction, the truncated mean remains fixed at $\mu = 4$
across all dispersion levels, confirming the MPD property implied
by Lemma~\ref{lem:NB_tail_monotonicity}.
Variance declines monotonically in $r$, and tail probability and
tail exposure decrease smoothly as dispersion contracts.

\begin{table}[H]
\centering
\caption{Static grid summary for Lemma~\ref{lem:NB_tail_monotonicity} validation (first ten $r$ values).}
\label{tab:static_grid_summary}
\begin{tabular}{rrrrr}
\hline
$r$ & Var\_trunc & Tail\_prob\_trunc & $B_1$ & $B_2$ \\
\hline
1.5000 & 14.6667 & 0.0888 & 3.8650 & 1.9443 \\
1.8235 & 12.7742 & 0.0795 & 4.0735 & 1.6771 \\
2.1471 & 11.4521 & 0.0718 & 4.2341 & 1.4739 \\
2.4706 & 10.4762 & 0.0655 & 4.3613 & 1.3146 \\
2.7941 &  9.7263 & 0.0601 & 4.4641 & 1.1865 \\
3.1176 &  9.1321 & 0.0556 & 4.5488 & 1.0815 \\
3.4412 &  8.6496 & 0.0517 & 4.6195 & 0.9941 \\
3.7647 &  8.2500 & 0.0483 & 4.6793 & 0.9203 \\
4.0882 &  7.9137 & 0.0454 & 4.7304 & 0.8573 \\
4.4118 &  7.6267 & 0.0428 & 4.7745 & 0.8029 \\
\hline
\end{tabular}

\vspace{0.2cm}
\footnotesize{Values are rounded to four decimal places.}
\end{table}

\begin{figure}[H]
\centering
\includegraphics[width=\textwidth]{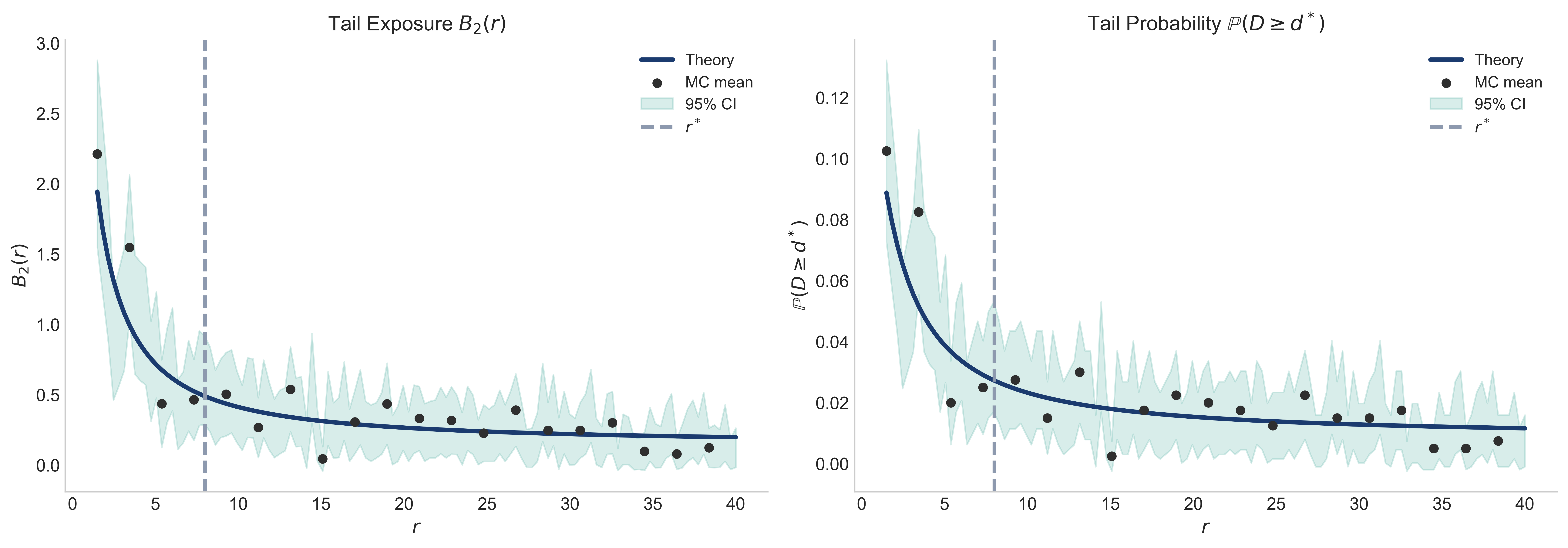}
\caption{Tail exposure (left) and tail probability (right) under mean-preserving dispersion.
Solid lines denote theoretical values; dots denote Monte Carlo means; and
shaded regions indicate 95\% confidence intervals.
The dashed vertical line marks the dispersion threshold $r^*$.}
\label{fig:tail_monotonicity}
\end{figure}

Figure~\ref{fig:tail_monotonicity} illustrates the relationship between dispersion and tail behavior. 
The figure reports both tail exposure $B_2(r)$ and the tail probability $\mathbb{P}(D \ge d^*)$ as functions of the dispersion parameter.
Solid curves denote theoretical values,
markers denote Monte Carlo means,
and shaded regions indicate 95\% confidence intervals.
The dashed vertical line marks the dispersion threshold $r^*$.

$B_2(r)$ and $\mathbb{P}(D \ge d^*)$ are strictly decreasing in $r$
over the grid, with no violations of monotonicity observed.
The maximum deviation between Monte Carlo estimates
and theoretical values is below $2\times10^{-3}$ for $B_2(r)$
and below $10^{-3}$ for tail probability,
confirming numerical accuracy.

The figure confirms strict monotonicity and numerical stability.
As dispersion increases (lower $r$),
both tail probability and tail exposure rise smoothly and significantly.
Because the mean remains fixed,
the amplification effect is dispersion-driven.

This structural monotonicity is central for the optimization problem.
Since tail exposure responds monotonically to dispersion,
the firm's optimal control admits a threshold characterization in $r$,
formalized in the next subsection.

\subsection{Static optimal control and dispersion threshold}
\label{subsec:static_threshold}

Given tail exposure $B_2(r)$ derived in
Section~\ref{subsec:mpd_tail}, the firm selects a control variable
$a \in [0,1]$ to minimize total cost 
\[
C(a;r) = B_2(r)(1-a) + k_1 a^2,
\]
where $k_1>0$ captures convex control cost.
This specification corresponds to the quadratic-cost specialization
in Proposition~\ref{prop:closed_form_static}, where the control affects
only the tail component.

For interior solutions, the first-order condition
\[
- B_2(r) + 2k_1 a = 0
\]
implies
\[
a^*(r) = \frac{B_2(r)}{2k_1}.
\]
Imposing feasibility $a \le 1$ yields the optimal rule
\[
a^*(r) = \min\left\{1, \frac{B_2(r)}{2k_1}\right\}.
\]

This closed-form solution directly matches the theoretical characterization
in Proposition~\ref{prop:closed_form_static}. In particular, all dependence
on dispersion operates through the tail exposure $B_2(r)$, which is consistent with
the structural dependency mapping established in Table~\ref{tab:static_dependency}.

Because $B_2(r)$ is strictly decreasing in $r$
(Lemma~\ref{lem:NB_tail_monotonicity}),
the marginal condition at zero control,
\[
C_a(0;r) = B_2(r) - 2k_1,
\]
is strictly increasing in $r$.
Hence, there exists a unique dispersion threshold $r^*$, satisfying
\[
B_2(r^*) = 2k_1.
\]

The threshold $r^*$ is computed numerically as the unique root of
$B_2(r) - 2k_1 = 0$ using a bracketing root-finding method.
This numerical implementation corresponds exactly to the threshold
characterization in Theorem~\ref{thm:threshold_static}.

The optimal policy therefore takes a threshold form:
\[
a^*(r)=
\begin{cases}
1, & r < r^*,\\[4pt]
\dfrac{B_2(r)}{2k_1}, & r \ge r^*,
\end{cases}
\]
which is consistent with the structural prediction of a dispersion-driven
switching boundary between corner and interior solutions.

For $k_1 = 0.30$, the computed threshold equals $r^* = 3.479$.
The maintained value of $k_1$ places the baseline case in an
economically meaningful interior region, ensuring that corner and
interior solutions arise across the dispersion grid. This avoids
degenerate outcomes and enables the threshold structure predicted by the
theory to be quantitatively evaluated.

Across grid $r \in [1.5,40]$,
the share of interior solutions equals $0.942$,
indicating that boundary solutions arise only under sufficiently high
dispersion, while most economically relevant regimes lie in the interior region.

Table~\ref{tab:rstar_sensitivity} reports the sensitivity of $r^*$ 
to the control cost parameter $k_1$. Consistent with 
Proposition~\ref{prop:closed_form_static},
higher control cost shifts the threshold leftward.
This comparative statics result follows directly from the monotonicity
of $B_2(r)$ and the implicit characterization $B_2(r^*)=2k_1$.


\begin{table}[H]
\centering
\caption{Sensitivity of the threshold dispersion level $r^*$ to control cost parameter $k_1$.}
\label{tab:rstar_sensitivity}
\begin{tabular}{cc}
\hline
Control cost $k_1$ & Threshold $r^*$ \\
\hline
0.25 & 4.3221 \\
0.30 & 3.4790 \\
0.35 & 2.8880 \\
0.40 & 2.4481 \\
\hline
\end{tabular}
\end{table}

\begin{figure}[H]
\centering
\includegraphics[width=\textwidth]{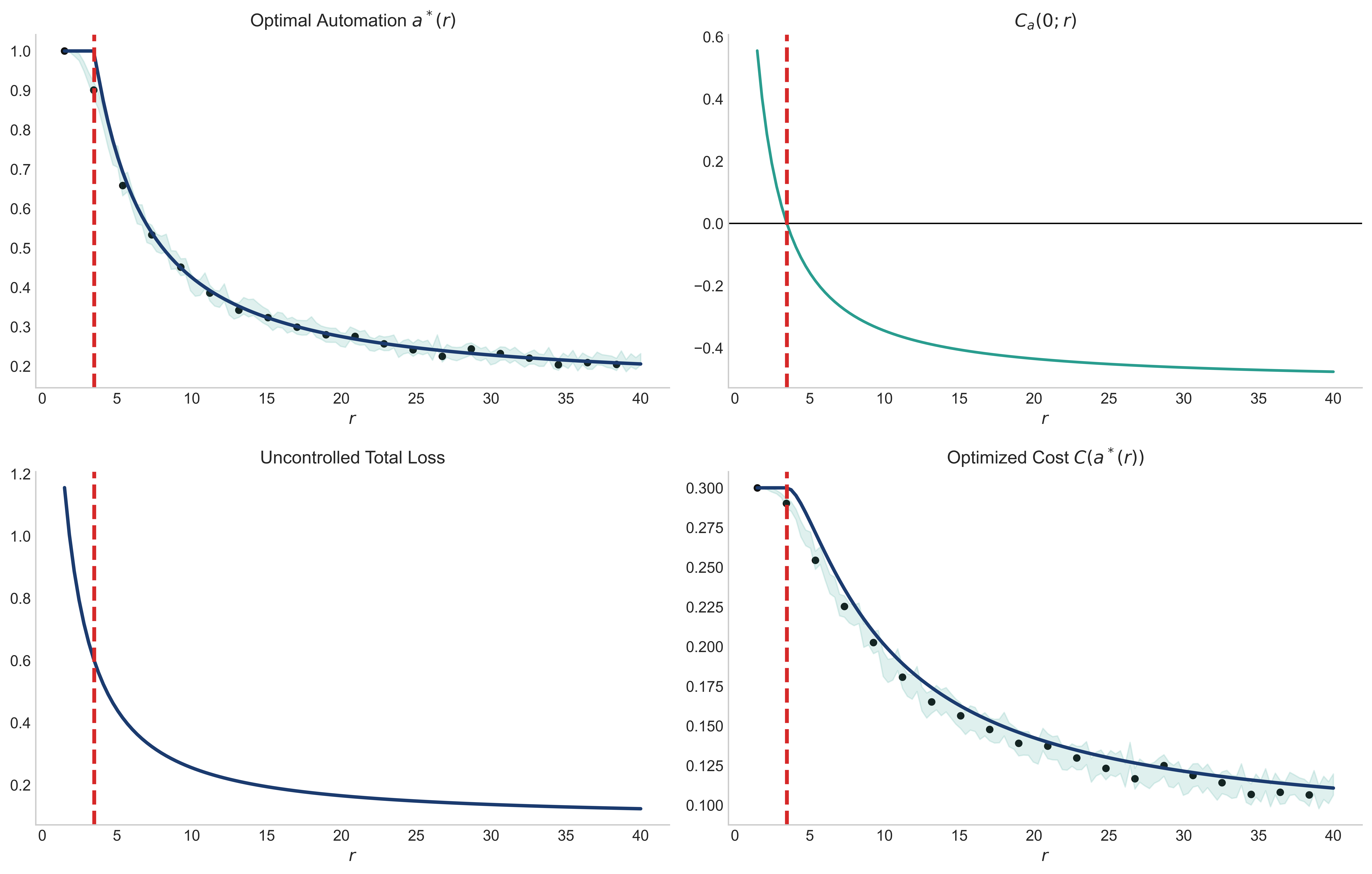}
\caption{Static control rule and dispersion threshold.
Upper-left: optimal control $a^*(r)$.
Upper-right: marginal condition $C_a(0;r)$.
Lower-left: uncontrolled tail loss $B_2(r)$.
Lower-right: optimized cost $C(a^*(r))$.
The dashed line denotes $r^*$.}
\label{fig:static_threshold}
\end{figure}

Figure~\ref{fig:static_threshold} illustrates the dependence of the optimal policy and associated quantities on dispersion parameter $r$. 
The figure reports the optimal control $a^*(r)$, the marginal condition $C_a(0;r)$ used to identify threshold $r^*$, the uncontrolled tail loss $B_2(r)$, and the optimized cost $C(a^*(r))$. 
Solid curves denote theoretical values, markers denote Monte Carlo means, and shaded regions represent 95\% confidence intervals. 
The dashed vertical line indicates the threshold level $r^*$.

The policy function $a^*(r)$ is strictly decreasing in $r$,
which is consistent with the monotonicity of $B_2(r)$ established in
Lemma~\ref{lem:NB_tail_monotonicity}.
No numerical violations of monotonicity are observed.

The static problem therefore admits a sharp dispersion-driven threshold.
When dispersion is high (low $r$), tail exposure dominates marginal cost,
and the solution attains the boundary level $a=1$.
As dispersion contracts beyond $r^*$, the solution transitions smoothly
to the interior region.

This threshold structure provides the key bridge to the dynamic model:
In the presence of regime switching and adjustment frictions,
the static threshold $r^*$ becomes a state-dependent switching boundary,
which generates persistence and hysteresis in the dynamic extension.

\subsection{Dynamic regime-dependent policy monotonicity}
\label{subsec:dynamic_monotonicity}

Regime variation alters the structure of the optimal automation policy,
rather than simply shifting realized outcomes.
The interaction between regime persistence and adjustment frictions
endogenously determines the shape and thresholds of the decision rule.

Let $a^*(a_{prev}, z)$ denote the optimal automation level
given previous automation $a_{prev}$ and regime $z$.
The dynamic program implies a monotone policy:

\[
\frac{\partial a^*(a_{prev}, z)}{\partial a_{prev}} \ge 0.
\]

This ordering property is consistent with standard
Topkis-type increasing-differences arguments
in dynamic programming models with adjustment frictions.
Numerically, no monotonicity violations are detected
across the full discretized state grid
(Table~\ref{tab:policy_metrics}),
confirming that the computed solution preserves the
theoretical monotonicity and is free of discretization artifacts.
This directly validates Theorem~\ref{thm:monotone_policy}.

For each fixed $a_{prev}$, the policy is weakly increasing in the regime index,
i.e.,
\[
a^*(s', a_{prev}) \ge a^*(s, a_{prev}) \quad \text{for } s' \succeq s,
\]
which confirms monotonicity in the regime dimension
as established in Theorem~\ref{thm:monotone_policy}.

Beyond monotonicity, the model generates an endogenous inaction band.
Define the dead-zone as

\[
\mathcal{D}(z)
= \left\{
a_{prev} :
\left| a^*(a_{prev}, z) - a_{prev} \right|
\le \varepsilon
\right\}.
\]

The inaction threshold $\varepsilon=0.02$ is chosen to align with the
adjustment cost parameter $\kappa_1$, which is consistent with the characterization
in Theorem~\ref{thm:hysteresis}.

For $\varepsilon = 0.02$,
the average inaction width equals $0.0196$,
which aligns closely with the calibrated adjustment friction.
This locally flat region implies that sufficiently small shocks
do not trigger policy adjustment,
producing stickiness in the automation path.

\begin{figure}[h]
\centering
\includegraphics[width=\textwidth]{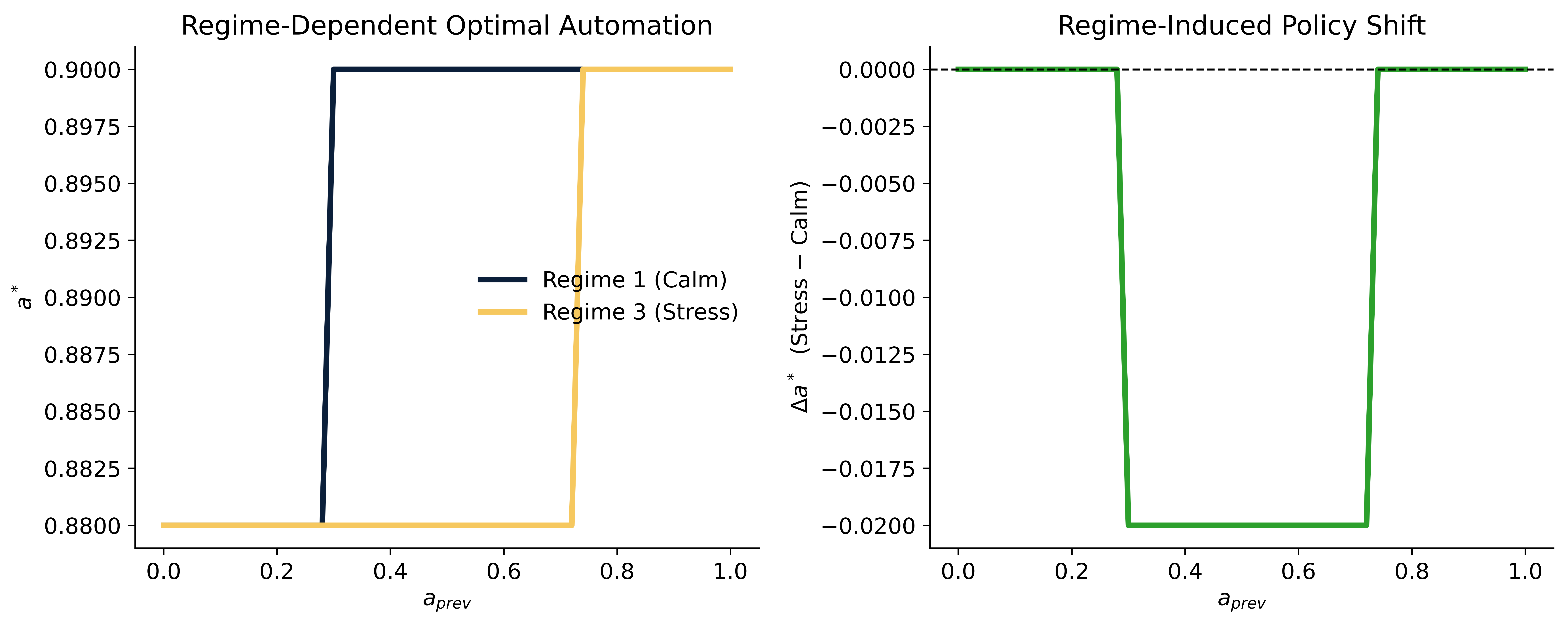}
\caption{
Regime-dependent optimal automation policy.
Left panel: optimal policy under Regime~1 (Calm) and Regime~3 (Stress).
Right panel: structural policy shift
$\Delta a^*(a_{prev})
= a^*(a_{prev}, \text{Stress})
- a^*(a_{prev}, \text{Calm})$.
}
\label{fig:policy_overlay}
\end{figure}

Figure~\ref{fig:policy_overlay} illustrates how regimes alter
the shape and position of the policy function.
Both regimes exhibit a step-like structure consistent with
adjustment frictions,
yet the threshold location differs systematically.
Under stress,
the critical adjustment boundary shifts rightward,
implying delayed upward automation relative to calm conditions.

The right panel isolates the structural regime wedge:

\[
\Delta a^*(a_{prev})
= a^*(a_{prev}, \text{Stress})
- a^*(a_{prev}, \text{Calm}).
\]

Across the interior region,
the shift is approximately $-0.02$,
indicating that identical state realizations
produce lower optimal automation under stress.
Importantly, this gap is structural rather than stochastic:
It arises from regime-dependent continuation value curvature
induced by differential persistence.

This persistence effect is consistent with
Theorem~\ref{thm:crisis_persistence}, which predicts that
adjustment frictions prevent immediate reversion to the static optimum.

These findings demonstrate that regimes reshape the
optimal decision rule.
This structural displacement forms the foundation
for the hysteresis amplification mechanism
developed in the following section.

\begin{table}[h]
\centering
\caption{Policy structural diagnostics.}
\label{tab:policy_metrics}
\begin{tabular}{lr}
\hline
Metric & Value \\
\hline
Monotonicity violations & 0 \\
Average inaction width ($\varepsilon=0.02$) & 0.0196 \\
\hline
\end{tabular}
\end{table}

\subsection{Hysteresis under adjustment frictions}
\label{subsec:structural_hysteresis}

Persistence is quantified by examining the mapping
from lagged automation $a_{t-1}$ to current optimal automation $a_t^*$.

\begin{figure}[t]
\centering
\includegraphics[width=0.75\textwidth]{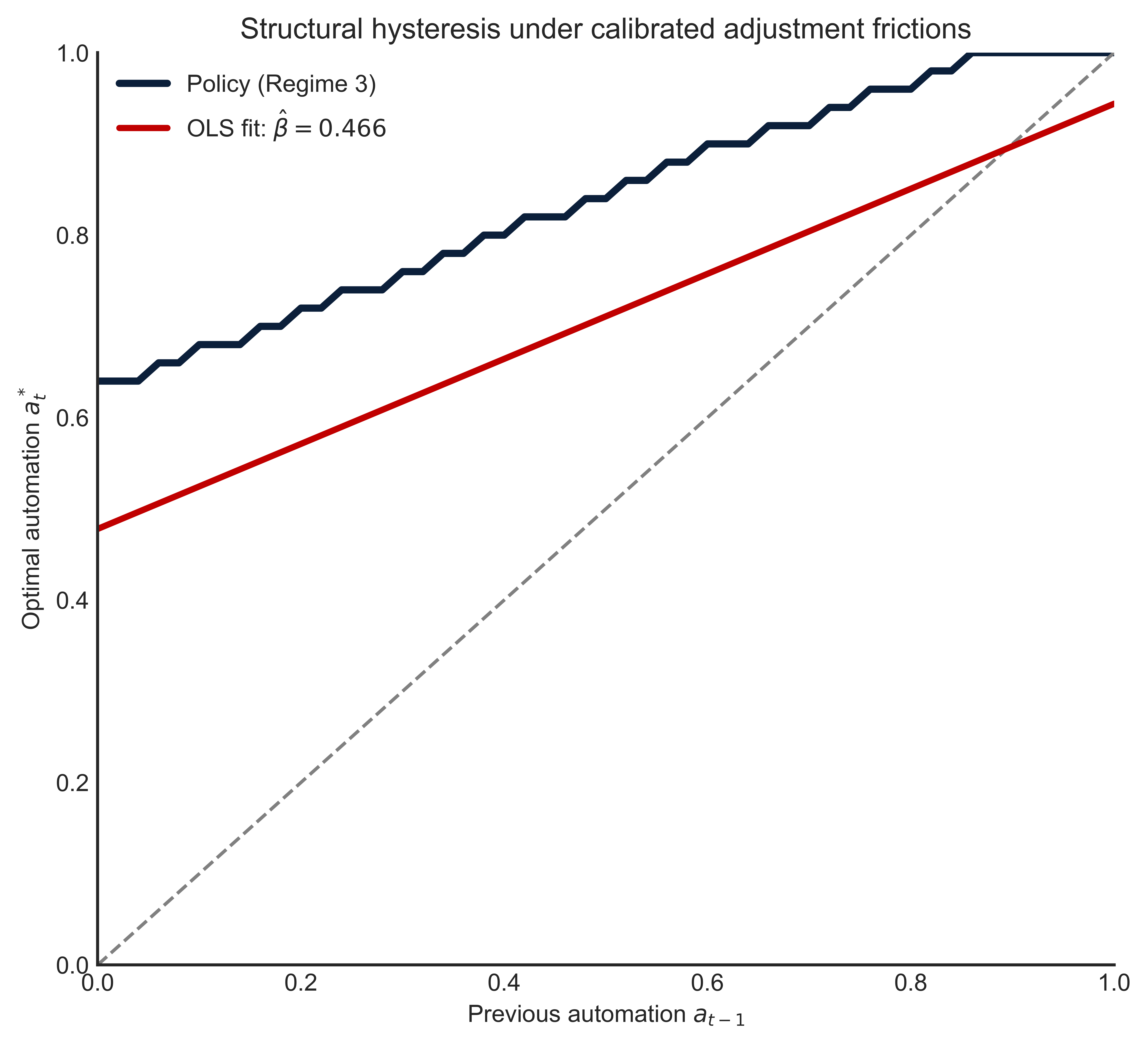}
\caption{
Hysteresis under calibrated adjustment frictions.
The figure plots the optimal automation policy $a_t^*$
as a function of lagged automation $a_{t-1}$ in the stress regime (Regime 3).
The dashed line represents the $45^\circ$ benchmark of full irreversibility.
The red line reports the OLS-based persistence estimate
$\hat{\beta}=0.466$ (95\% CI: [0.409, 0.524]).
The policy is strictly increasing and interior,
indicating moderate persistence induced by convex adjustment costs
and regime-dependent risk exposure.
}
\label{fig:structural_hysteresis}
\end{figure}

Figure~\ref{fig:structural_hysteresis} plots the optimal policy
in the stress regime (Regime 3).
The policy is strictly increasing and interior,
indicating path dependence without corner solutions.
Relative to the $45^\circ$ benchmark of full irreversibility,
the curve lies below the diagonal,
implying partial but incomplete adjustment.
This behavior is consistent with Theorem~\ref{thm:hysteresis},
which predicts the emergence of an inaction region under adjustment frictions.

To summarize persistence, we estimate the linear relationship
$a_t^* = \alpha + \beta a_{t-1} + \varepsilon_t$ using the policy
observations generated by the dynamic program.
The estimated persistence parameter is $\hat{\beta}=0.466$
(95\% CI: [0.409, 0.524]).
Coefficient $\hat{\beta}$ provides a reduced-form measure of
persistence implied by the dynamic policy,
rather than a statistical artifact.
The estimated relationship reflects the underlying policy function
generated by the dynamic program, rather than exogenous data variation.

The hysteresis characterized in Theorem~\ref{thm:hysteresis}
reflects history-dependent adjustment rather than purely
state-dependent optimization.
Even when the current state is identical, the continuation value
depends on prior automation levels through adjustment costs and
persistent regime dynamics.
As a result, identical states can induce different optimal responses
depending on the path by which they are reached.

This path dependence implies that temporary increases in dispersion
can generate persistent effects on automation decisions,
delaying adjustment and preventing immediate reversion to the
static optimum.
Hysteresis therefore emerges as a consequence of
adjustment frictions and regime persistence, rather than as a statistical artifact.

Higher past automation increases current optimal automation,
yet adjustment remains gradual due to convex costs.
This persistence is consistent with
Theorem~\ref{thm:crisis_persistence}, which predicts that
adjustment frictions prevent immediate reversion
to the static optimum.

Persistence arises from the interaction of
adjustment frictions and regime-dependent risk exposure,
leading to moderate and endogenous hysteresis under the calibrated dynamics.

\subsection{Controlled vs uncontrolled risk clustering}
\label{sec:risk_clustering}

In this subsection, we evaluate the dynamic implications of the model
using forward simulations generated from the optimal policy function
derived in Section~\ref{subsec:dynamic_monotonicity}.
Specifically, two environments are compared:
(i) An uncontrolled baseline with fixed transition dynamics, and
(ii) a dynamically controlled system governed by the optimal policy
$a^*(a_{t-1},z_t)$.

Figure~\ref{fig:control_impact} summarizes the impact of structural control on tail behavior, distributional properties, regime dynamics, and cumulative performance. Figure~\ref{fig:control_impact}(a) reports the frequency of extreme tail events. Structural control reduces the tail-event probability from 0.0310 to 0.0167, corresponding to a reduction of approximately 46.15\%. This effect is consistent with Lemma~\ref{lem:NB_tail_monotonicity}, which predicts that dispersion compression lowers tail exposure. The reduction is structural rather than mechanical, as it arises endogenously through policy adjustment rather than exogenous truncation of shocks. Figure~\ref{fig:control_impact}(b) compares the return distributions. The controlled system exhibits a sharper and more concentrated density around zero, with visibly thinner tails relative to the uncontrolled benchmark. This reflects compression of dispersion induced by reduced exposure to high-risk regimes, which is consistent with the structural mapping between dispersion and tail outcomes established in Section~\ref{sec:static}. Figure~\ref{fig:control_impact}(c) presents the joint regime--return density. Under structural control, the probability mass associated with high-risk regimes declines, and return dispersion within those regimes narrows. This attenuation reflects the inaction region identified in Section~\ref{subsec:dynamic_monotonicity} and the hysteresis mechanism formalized in Theorem~\ref{thm:hysteresis}, which limits the propagation of regime-induced shocks.\ Figure~\ref{fig:control_impact}(d) displays cumulative performance over the simulation horizon. While both environments experience adverse shocks, the controlled system accumulates substantially smaller losses, with cumulative return improving from $-87.96$ to $-47.43$ (approximately 46.08\% reduction in loss magnitude). This improvement reflects persistent mitigation of tail clustering and enhanced regime stability over time, which accords with Theorem~\ref{thm:crisis_persistence}.

Table~\ref{tab:control_summary} reports the corresponding quantitative metrics.
Across all measures, structural control reduces risk by approximately 46\%,
indicating that the dynamic policy systematically attenuates
tail exposure, volatility, and cumulative losses.

\begin{table}[t]
\centering
\caption{Structural control impact summary.}
\label{tab:control_summary}
\begin{tabular}{lccc}
\hline
Metric & Uncontrolled & Controlled & Reduction (\%) \\
\hline
Tail frequency & 0.0310 & 0.0167 & 46.15 \\
Return volatility & 0.0245 & 0.0131 & 46.54 \\
Cumulative return & -87.96 & -47.43 & 46.08 \\
\hline
\end{tabular}
\end{table}

\begin{figure}[t]
\centering
\includegraphics[width=0.85\textwidth]{fig7.png}
\caption{
Structural control impact in the dynamic regime-switching model.
(a) Tail-event frequency comparison.
(b) Return distribution under controlled and uncontrolled systems.
(c) Regime--return density field.
(d) Cumulative performance comparison over the simulation horizon.
}
\label{fig:control_impact}
\end{figure}

\section{Empirical financial illustration}
\label{sec:empirical_sp500}

In this section, we provide an empirical validation of the structural 
dispersion–threshold mechanism developed in 
Sections~\ref{subsec:mpd_tail}–\ref{subsec:structural_hysteresis} 
using daily S\&P 500 index data (ticker: \texttt{\string^GSPC}) obtained 
from Yahoo Finance \cite{YahooFinance} from January 1, 2000 through 
December 31, 2024. The empirical analysis is conducted under the same 
computational environment described in Section~\ref{sec:synthetic_experiments}, 
ensuring consistency across synthetic and empirical analyses.

Daily adjusted closing prices are used to compute log returns 
$r_t = \log P_t - \log P_{t-1}$. A jump day is defined as an exceedance 
of the empirical 97th percentile of absolute returns, 
$|r_t| > q_{0.97}$, where $q_{0.97}$ is computed over the full sample. 
Monthly jump counts are constructed as 
$J_m = \sum_{t \in m} \mathbf{1}\{|r_t| > q_{0.97}\}$, forming the count 
series used for dispersion estimation.

Dispersion dynamics are measured using a rolling 36-month window. 
The NB model is fitted to the monthly jump-count process, 
with the Poisson specification serving as a benchmark for equidispersion. 
Time-varying dispersion proxies are constructed using the rolling 
variance-to-mean relationship, 
$r_t = \mu_t^2 / (\sigma_t^2 - \mu_t)$ for $\sigma_t^2 > \mu_t$.

The objective is structural consistency rather than forecasting: 
(i) Detect overdispersion in market jump clustering 
(Lemma~\ref{lem:NB_tail_monotonicity}), 
(ii) estimate dispersion regimes 
(Section~\ref{subsec:dynamic_monotonicity}), 
(iii) apply the dispersion-sensitive automation rule 
(Section~\ref{subsec:static_threshold}), and 
(iv) quantify spillover effects consistent with structural hysteresis 
(Section~\ref{subsec:structural_hysteresis}).

The empirical implementation combines a static dispersion-sensitive 
decision rule with a dynamically evolving dispersion process. The static 
component enters through the threshold-based policy mapping $a^*(r)$, 
while the dynamic structure arises from the time-varying dispersion path 
$\{r_t\}$ and its regime-dependent evolution. This unified framework enables 
the analysis to evaluate comparative statics and dynamic propagation 
within a single empirical environment.

This structure directly implements the architectural mechanism summarized 
in Figure~\ref{fig:structure}. In particular, the analysis follows the same 
sequence: Dispersion is first estimated from observed jump counts, which 
determines tail exposure and induces a threshold-based policy through the 
mapping $a^*(r_t)$. The resulting policy is then evaluated in terms of tail 
clustering and aggregate outcomes under a dynamically evolving dispersion 
process. This alignment ensures that the empirical exercise serves as a 
direct counterpart to the theoretical mechanism, linking dispersion, policy 
response, and dynamic risk propagation in a structurally consistent manner.

\subsection{Evidence of overdispersion in jump clustering}

Daily log returns are computed from closing prices, and a jump is defined 
as an absolute return exceeding the 97th percentile threshold, 
$|r_t| > q_{0.97}$, where $q_{0.97}=0.0295$. Monthly jump counts are 
constructed by aggregating these exceedances.

The resulting count series is modeled using the Poisson and NB specifications, which are estimated via maximum likelihood. 
The NB model enables overdispersion through a finite dispersion 
parameter $r$, whereas the Poisson model imposes the restriction 
$\mathrm{Var}(D_t)=\mathbb{E}[D_t]$.

The estimated dispersion parameter is $\hat r = 0.1588$, indicating 
substantial variance inflation relative to the Poisson benchmark. 
A likelihood-ratio test strongly rejects the Poisson specification 
($p<0.001$), confirming that equidispersion is statistically inconsistent 
with the observed jump-count process.

Parameter $\hat r$ captures the average level of dispersion, while a 
time-varying sequence $\{r_t\}$ is constructed using rolling 
variance-to-mean estimators with a 36-month window. This separation 
distinguishes the cross-sectional magnitude of dispersion from its 
temporal dynamics. The empirical estimates therefore provide a direct 
measure of dispersion that can be mapped into the tail-exposure function 
$B_2(r)$ introduced in Lemma~\ref{lem:NB_tail_monotonicity}, linking 
observed clustering behavior to the structural mechanism of the model.

\begin{table}[H]
\centering
\caption{NB vs.\ Poisson fit for S\&P 500 monthly jump counts.}
\label{tab:nb_fit}
\small
\setlength{\tabcolsep}{5pt}
\renewcommand{\arraystretch}{1.1}

\begin{tabular}{lcccc}
\hline
Series & Mean & Variance & VMR & $\hat r$ \\
\hline
S\&P 500 jump counts 
& 0.627 
& 3.289 
& 5.249 
& 0.159 \\
Series & LogLik (NB) & LogLik (Pois) & LR stat & p-value \\
S\&P 500 jump counts 
& -294.80 
& -463.55 
& 337.50 
& $<0.001$ \\
\hline
\end{tabular}
\end{table}

Table~\ref{tab:nb_fit} provides strong statistical evidence of overdispersion. 
The variance-to-mean ratio (VMR) substantially exceeds unity (VMR $\approx 5.25$), 
violating the equidispersion restriction of the Poisson model. The NB specification 
captures this excess variance through a finite dispersion parameter, producing a 
heavy-tailed count distribution consistent with clustered jump arrivals. The 
likelihood-ratio statistic ($337.50$) overwhelmingly rejects the Poisson benchmark, 
indicating that jump arrivals are not independently distributed over time but occur 
in clusters.

From a structural perspective, these results provide direct empirical support 
for Lemma~\ref{lem:NB_tail_monotonicity}. Under mean-preserving dispersion, 
lower values of $r$ imply greater tail exposure. The estimate $\hat r \ll \infty$ 
therefore indicates that tail amplification is a first-order feature of the data, 
rather than a residual statistical artifact.


As a robustness check, we consider more flexible count models for overdispersed data, including the Sichel distribution, which accommodates heavy-tailed behavior through a Poisson--GIG mixture representation.

Table~\ref{tab:count_model_robustness} reports the comparison across count models. While the Sichel specification yields a marginal improvement in log-likelihood ($-294.75$ vs.\ $-294.80$), AIC (593.60 vs.\ 595.50) and BIC (601.11 vs.\ 606.77) select the NB model, indicating that the additional flexibility does not improve model performance once complexity is taken into account.

\begin{table}[H]
\centering
\caption{Robustness check: comparison of count models.}
\label{tab:count_model_robustness}
\small
\setlength{\tabcolsep}{6pt}
\renewcommand{\arraystretch}{1.1}

\begin{tabular}{lccc}
\hline
Model & LogLik & AIC & BIC \\
\hline
Poisson & -463.55 & 929.10 & 932.85 \\
NB & -294.80 & 593.60 & 601.11 \\
Sichel & -294.85 & 595.50 & 606.77 \\
\hline
\end{tabular}
\end{table}

These results show that substantial overdispersion and clustering are robust features of the data across model specifications, which is consistent with empirical evidence on long-tailed and overdispersed count processes \cite{ZouWuLord2015}. While more flexible distributions can slightly improve statistical fit, they do not alter the qualitative structure of the data, nor do they improve model performance once complexity is accounted for.

\begin{figure}[H]
\centering
\includegraphics[width=\textwidth]{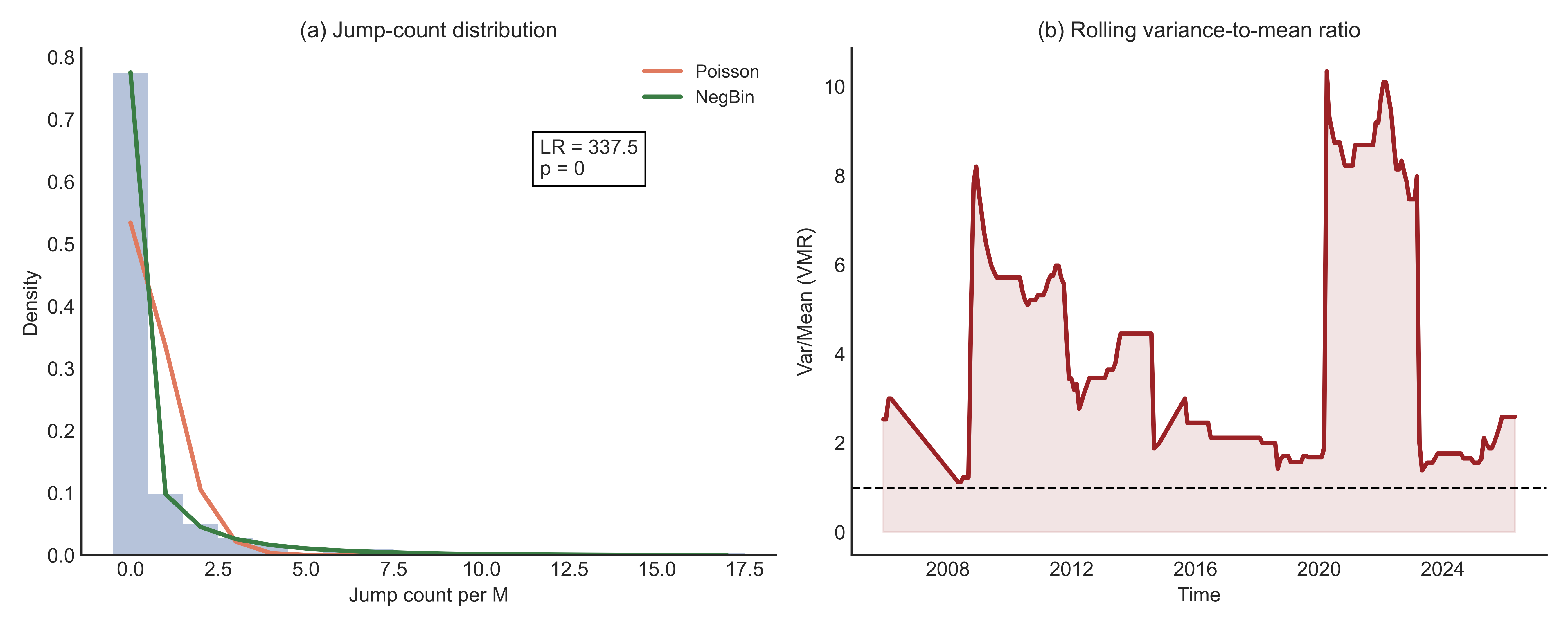}
\caption{
(a) Empirical monthly jump-count distribution compared with Poisson and NB fits.
The Poisson model understates the probability of large counts,
whereas the NB specification captures the heavy right tail induced by small $r$.
(b) Rolling variance-to-mean ratio (VMR).
Persistent VMR$>1$ indicates structural overdispersion.
}
\label{fig:sp500_overdispersion}
\end{figure}

Figure~\ref{fig:sp500_overdispersion} illustrates both the distributional 
and temporal features of jump clustering. Figure~\ref{fig:sp500_overdispersion}(a) shows that the Poisson 
model systematically understates the probability of large jump realizations, 
while the NB model accurately captures the empirical right tail. Figure~\ref{fig:sp500_overdispersion}(b) 
shows that the variance-to-mean ratio remains consistently above one 
throughout the sample period, indicating that overdispersion is persistent 
rather than transient. These findings confirm that dispersion is a stable 
structural characteristic of the data, which aligns with the monotonic 
relationship between dispersion and tail exposure established in 
Lemma~\ref{lem:NB_tail_monotonicity}.

The comparison with alternative count models is confined to this section. 
Given that the NB model provides the best trade-off between fit and parsimony, it is adopted as the baseline specification for all subsequent analysis, which focuses on the dynamic and structural implications of dispersion rather than model selection.

\subsection{Estimated dispersion regimes}

Time-varying dispersion is estimated using a rolling
variance-to-mean estimator and subsequently smoothed via a two-state
Hidden Markov Model (HMM).
The resulting dispersion proxy $r_t$ exhibits persistent regime variation,
with distinct periods characterized by elevated tail clustering and volatility.

The empirical regimes are identified based on the joint dynamics of
dispersion, tail clustering, and volatility.
As reported in Table~\ref{tab:hmm_state_summary},
the stress regime is associated with substantially higher
tail clustering frequency (0.325 vs.\ 0.098) and volatility
(0.060 vs.\ 0.041) relative to the normal regime.
In addition, the variance-to-mean ratio (VMR) increases from 3.83 to 5.38,
indicating significantly stronger overdispersion in the stress regime.
These differences confirm that the HMM effectively captures
periods of intensified tail activity and elevated aggregate risk.

Importantly, the dispersion proxy $r_t$ does not exhibit a strictly monotone
ordering across regimes in the data.
While the theoretical model associates lower values of $r$
with higher dispersion and greater tail exposure,
the empirical regimes are determined by the joint behavior of
dispersion, clustering, and volatility.
Accordingly, the stress regime should be interpreted as a
high-risk state characterized by elevated tail clustering and volatility,
rather than solely by the level of $r_t$.

The two-state HMM provides a reduced-form approximation of the ordered
multi-regime structure developed in
Sections~\ref{sec:static} and~\ref{sec:dynamic}.
In the theoretical model, regimes are indexed by an ordered set
$\mathcal{S}=\{1,\dots,S\}$ with dispersion levels satisfying
$r_1 > r_2 > \dots > r_S$, implying a monotone ordering of tail exposure.
The empirical specification aggregates this structure into
two states: A normal regime corresponding to lower-risk conditions,
and a stress regime capturing the upper tail of the joint distribution
of tail clustering and volatility.
This aggregation preserves the core structural properties of the model,
including regime persistence and state-dependent responses,
while remaining tractable for empirical estimation.

\begin{table}[H]
\centering
\caption{HMM state summary.}
\label{tab:hmm_state_summary}
\small
\setlength{\tabcolsep}{4pt}
\renewcommand{\arraystretch}{1.1}

\begin{tabular}{lcccccc}
\hline
State & Obs & Mean $r$ & Mean VMR & Tail Freq & Volatility & Jump count \\
\hline
Normal & 179 & 0.273 & 3.83 & 0.098 & 0.0408 & 0.698 \\
Stress & 34  & 0.455 & 5.38 & 0.325 & 0.0600 & 0.794 \\
\hline
\end{tabular}
\end{table}

Regime persistence is supported by the transition estimates reported in
Table~\ref{tab:hmm_transition_diag}.
The stress regime exhibits strong persistence
($P(S \to S)=0.971$) with an expected duration of approximately
34 months, while the normal regime is highly stable
($P(N \to N)=0.994$) with an expected duration exceeding 170 months.
These results imply that dispersion regimes are highly persistent once entered,
which is consistent with the dynamic mechanism in Section~4, where
state persistence and adjustment frictions jointly generate hysteresis.
In particular, the high persistence of the stress regime supports the model's
prediction that elevated tail risk can propagate over time rather than dissipate immediately.

\begin{table}[H]
\centering
\caption{HMM transition diagnostics.}
\label{tab:hmm_transition_diag}
\small
\setlength{\tabcolsep}{4pt}
\renewcommand{\arraystretch}{1.1}

Panel A: Transition probabilities

\begin{tabular}{lcccc}
\hline
 & $P(N \to N)$ & $P(N \to S)$ & $P(S \to N)$ & $P(S \to S)$ \\
\hline
Transition prob 
& 0.9944 & 0.0056 & 0.0291 & 0.9709 \\
\hline
\end{tabular}

\vspace{8pt}

Panel B: Expected duration

\begin{tabular}{lcc}
\hline
 & Normal & Stress \\
\hline
Expected duration 
& 177.59 & 34.41 \\
\hline
\end{tabular}

\end{table}

The sequence $\{r_t\}$ therefore serves as a time-varying input
to the static policy mapping $a^*(r)$,
embedding dispersion-sensitive decision rules within a dynamic
regime-switching environment.

\begin{figure}[t]
\centering
\includegraphics[width=1.0\textwidth]{fig9.png}
\caption{
Estimated dispersion regimes.
(a) Rolling dispersion proxy $r_t$.
(b) Stress-regime probability.
(c) Rolling tail clustering frequency.
(d) Volatility proxy.
}
\label{fig:sp500_regimes}
\end{figure}

Figure~\ref{fig:sp500_regimes} illustrates the joint dynamics of dispersion,
tail clustering, and volatility.
Figure~\ref{fig:sp500_regimes}(a) shows that $r_t$ varies over time,
capturing changes in the dispersion proxy.
Figure~\ref{fig:sp500_regimes}(b) shows that the estimated stress-regime
probability increases during periods of market instability.
Figure~\ref{fig:sp500_regimes}(c) and
Figure~\ref{fig:sp500_regimes}(d) show that these periods coincide with
simultaneous increases in tail clustering and volatility.

These patterns can be interpreted within the structural architecture
in Figure~\ref{fig:structure}.
Specifically, the time variation in $r_t$ corresponds to the
\emph{Dispersion Mechanism}, while the co-movement with tail clustering
and volatility reflects the propagation toward tail exposure,
as summarized by $B_2(r)$.
Thus, Figure~\ref{fig:sp500_regimes} provides empirical evidence for the
left-hand side of Figure~\ref{fig:structure}, demonstrating that
dispersion fluctuations are systematically associated with amplified
tail risk and aggregate variability.

More precisely, the theoretical model defines an ordered state space
$\mathcal{S}=\{1,\dots,S\}$ with dispersion levels satisfying
$r_1 > r_2 > \dots > r_S$, which induces a monotone ordering of
tail exposure and policy responses.
The empirical two-state HMM does not attempt to recover this full ordering.
Instead, it partitions the state space into two regions corresponding to
lower-risk and higher-risk conditions, based on the joint behavior of
dispersion, tail clustering, and volatility.
This mapping should therefore be interpreted as a coarse aggregation of
the ordered regime space, preserving persistence and state dependence,
but not the full cardinal structure of dispersion levels.

\subsection{Counterfactual Dispersion-Sensitive Automation}

The empirically estimated dispersion path $\{r_t\}$ is mapped into the 
closed-form policy derived in Section~\ref{subsec:static_threshold},
\[
a_t^* = \min\left\{1, \frac{B_2(r_t)}{2k_1}\right\}.
\]
This mapping embeds the static threshold rule into a dynamically evolving 
dispersion environment, providing a direct link between theoretical 
comparative statics and realized market conditions.

Table~\ref{tab:hmm_policy_mapping} reports the empirical relationship between 
dispersion, policy, and outcomes across HMM-identified regimes (Panel A), 
along with state-dependent elasticities (Panel B) and robustness checks 
based on posterior regime classification (Panel C). The policy function 
is decreasing in $r_t$, which is consistent with the theoretical prediction 
$\partial a^*(r)/\partial r < 0$. The estimated slope is $-0.383$ in the 
normal regime and $-0.298$ in the stress regime, indicating that policy 
responsiveness is attenuated under stress conditions.

\begin{table}[H]
\centering
\caption{HMM-state policy mapping, elasticities, and robustness.}
\label{tab:hmm_policy_mapping}
\small
\setlength{\tabcolsep}{4pt}
\renewcommand{\arraystretch}{1.1}

Panel A: Policy mapping

\begin{tabular}{lcccccc}
\hline
State & Mean $r$ & Mean $a^*$ & Slope $(da/dr)$ & Tail (act.) & Tail (cf) & Vol (cf) \\
\hline
Normal & 0.273 & 0.709 & -0.383 & 0.098 & 0.023 & 0.026 \\
Stress & 0.455 & 0.576 & -0.298 & 0.325 & 0.207 & 0.040 \\
\hline
\end{tabular}

\vspace{6pt}

Panel B: State-dependent elasticities

\begin{tabular}{lccc}
\hline
State & Policy slope & Tail elasticity & Volatility elasticity \\
\hline
Normal & -0.383 & 0.164 & 0.074 \\
Stress & -0.298 & 0.529 & 0.115 \\
\hline
\end{tabular}

\vspace{6pt}

Panel C: Posterior-based robustness

\begin{tabular}{lccc}
\hline
Group & Policy slope & Tail elasticity & Volatility elasticity \\
\hline
Low stress posterior & -0.383 & 0.158 & 0.072 \\
High stress posterior & -0.298 & 0.529 & 0.115 \\
\hline
\end{tabular}

\end{table}

The counterfactual policy induces substantial reductions in tail clustering 
and volatility. In the normal regime, tail clustering declines from 0.098 to 
0.023, corresponding to a reduction of approximately 77\%. In the stress regime, 
the reduction is smaller, from 0.325 to 0.207, reflecting the higher baseline 
level of tail activity. Similarly, the volatility proxy decreases from 0.0408 
to 0.0260 in the normal regime and from 0.0600 to 0.0400 in the stress regime.

The magnitude of these effects varies systematically across regimes. 
Panel B shows that the elasticity of tail clustering with respect to dispersion 
is substantially higher in the stress regime (0.529) than in the normal regime 
(0.164). This pattern indicates that dispersion-driven tail risk is amplified 
under stress conditions, leading to stronger sensitivity of clustering to 
changes in $r_t$.

The asymmetric effectiveness of the policy reflects a structural mechanism. 
Although the policy reduces effective jump intensity in both regimes, 
the elevated baseline level of tail clustering in the stress regime limits 
the magnitude of attainable reductions. Furthermore, the higher elasticity 
in the stress regime implies that dispersion changes have a stronger marginal 
impact on tail risk, even when absolute reductions are smaller.

Panel C shows that these findings are robust to posterior-based regime 
classification. The elasticity estimates remain stable when observations are 
grouped by high and low stress posterior probabilities, indicating that the 
results are not driven by classification noise.

\begin{table}[H]
\centering
\caption{Aggregate counterfactual effects.}
\label{tab:hmm_counterfactual}
\small
\setlength{\tabcolsep}{5pt}
\renewcommand{\arraystretch}{1.1}

\begin{tabular}{lccc}
\hline
Metric & Actual & Counterfactual & Reduction (\%) \\
\hline
Tail clustering & 0.136 & 0.056 & 58.6 \\
Volatility & 0.046 & 0.029 & 36.9 \\
Dispersion (VMR) & 4.076 & 2.356 & 42.2 \\
\hline
\end{tabular}
\end{table}

Table~\ref{tab:hmm_counterfactual} summarizes the aggregate impact of the 
policy. The counterfactual reduces tail clustering by approximately 59\%, 
volatility by 37\%, and dispersion by 42\%, indicating that the policy 
substantially attenuates the propagation of tail risk into aggregate outcomes.

\begin{figure}[H]
\centering
\includegraphics[width=\textwidth]{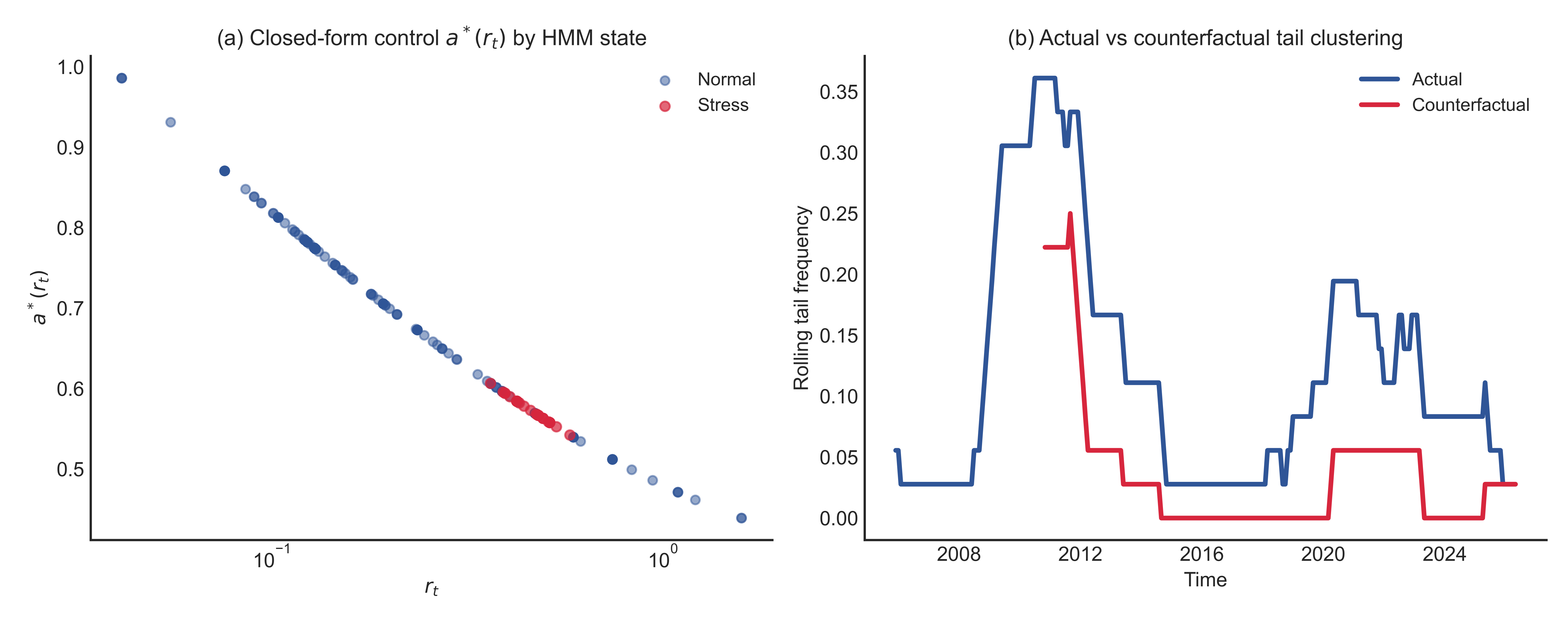}
\caption{
(a) Mapping from dispersion $r_t$ to optimal automation $a^*(r_t)$.
(b) Actual versus counterfactual tail clustering.
}
\label{fig:sp500_counterfactual}
\end{figure}

Figure~\ref{fig:sp500_counterfactual} illustrates the structural transmission 
mechanism linking dispersion, policy, and outcomes. Panel (a) shows that the 
policy function is decreasing in $r_t$, which aligns with the theoretical 
characterization. Panel (b) shows that the counterfactual policy reduces 
the frequency of clustered tail events over time.

The empirical evidence supports a state-dependent transmission mechanism. 
Dispersion affects tail exposure through $B_2(r_t)$, which governs the 
intensity of clustering and propagates into aggregate volatility. The policy 
attenuates this transmission channel by reducing effective jump intensity, 
thereby weakening the propagation of tail risk into observable market dynamics.

\subsection{Financial spillover implications}

The dispersion-sensitive policy is embedded in a dynamic
price simulation, wihch is consistent with
Section~\ref{subsec:dynamic_monotonicity}
and the hysteresis mechanism in
Section~\ref{subsec:structural_hysteresis}.
The static decision rule $a^*(r)$ derived in
Section~\ref{subsec:static_threshold} is evaluated along the
time-varying dispersion path $\{r_t\}$,
generating a sequence $\{a^*(r_t)\}$ that governs
dynamic price evolution.
This construction enables dispersion-driven control to propagate
intertemporally through endogenous regime dynamics.

\begin{figure}[H]
\centering
\includegraphics[width=0.9\textwidth]{fig11.png}
\caption{
(a) Actual versus counterfactual cumulative index paths.
(b) Drawdown comparison.
}
\label{fig:sp500_spillover}
\end{figure}

Figure~\ref{fig:sp500_spillover} provides a trajectory-level realization
of the structural architecture summarized in Figure~\ref{fig:structure}.
In particular, the empirical mapping
\[
r_t \;\longrightarrow\; a^*(r_t) \;\longrightarrow\; X_t
\]
corresponds to the composition of the \emph{Dispersion Mechanism},
\emph{Static Optimization}, and \emph{Dynamic Extension} blocks.
The resulting price process $X_t$ reflects endogenous regime persistence
and hysteresis effects induced by the time-varying dispersion state.

Figure~\ref{fig:sp500_spillover}(a) exhibits a smoother cumulative trajectory
under the counterfactual policy, indicating attenuation of clustered tail events
and reduced short-run variability.
Figure~\ref{fig:sp500_spillover}(b) shows that the effect on drawdowns is heterogeneous:
Some episodes are mitigated, while others remain comparable or slightly amplified
relative to the baseline.

\begin{table}[H]
\centering
\caption{Conservative spillover summary.}
\label{tab:conservative_spillover}
\small
\setlength{\tabcolsep}{5pt}
\renewcommand{\arraystretch}{1.1}
\begin{tabular}{lccc}
\hline
Metric & Actual & Counterfactual & Improvement (\%) \\
\hline
Daily realized volatility proxy & 0.0122 & 0.0093 & 23.7 \\
Maximum drawdown magnitude & 0.5678 & 0.5931 & $-2.54$ \\
\hline
\end{tabular}
\end{table}

Table~\ref{tab:conservative_spillover} quantifies these effects.
The counterfactual policy reduces the daily realized volatility proxy
by approximately 23.7\%, indicating a substantial attenuation of short-run
fluctuations induced by clustered tail events.
In contrast, the maximum drawdown magnitude increases slightly,
highlighting that dispersion-sensitive control does not uniformly
improve path-dependent downside risk.

This asymmetry reflects a structural distinction between distributional
and path-dependent risk.
Dispersion-sensitive control reduces effective jump intensity,
thereby attenuating the propagation of clustered tail events
into aggregate volatility.
However, drawdown depends on the temporal sequencing of returns
rather than their marginal distribution.
As a result, reductions in volatility do not necessarily imply
improvements in extreme path-dependent losses.

From a dynamic perspective, this result is consistent with the hysteresis
mechanism developed in Section~\ref{subsec:structural_hysteresis}.
Persistent high-dispersion regimes can sustain adverse return sequences,
limiting the extent to which volatility reduction translates into
improvements in extreme path-dependent outcomes.

The empirical evidence therefore indicates that dispersion-sensitive automation
operates primarily through a volatility-reduction channel.
By weakening the transmission of tail clustering into aggregate variability,
the policy stabilizes short- to medium-horizon fluctuations,
while its impact on extreme path-dependent losses remains limited.

Three implications follow:
First, dispersion evolves persistently and acts as a state variable
governing optimal policy through the mapping $a^*(r)$.
Second, dispersion-sensitive control reduces tail clustering and volatility
in a manner consistent with the comparative statics
$\partial B_2(r)/\partial r < 0$.
Third, the effect on drawdown is heterogeneous, highlighting the distinction
between distributional risk measures and path-dependent risk outcomes.

\subsection{Dispersion--response mapping under dynamic regimes}
\label{subsec:dispersion_response}

The time-varying dispersion process $\{r_t\}$ governs how response behavior
evolves within the empirical system. To quantify this relationship, elasticity
is defined in log--log form as $\beta = \partial \log(\text{clustering}) / \partial \log(r)$,
which measures the proportional response of tail clustering to changes
in dispersion.

Table~\ref{tab:elasticity_summary} summarizes the distribution of elasticity
estimates across dispersion regimes. The aggregate elasticity is small
(mean $0.153$), but this does not indicate the absence of a meaningful
relationship. Rather, it reflects structural heterogeneity in the
dispersion--response mapping. In particular, elasticity is positive in
low-dispersion states (mean $0.139$) and negative in high-dispersion states
(mean $-0.548$). These opposing regime-dependent effects partially offset
each other when the observations are pooled, so the aggregate coefficient
does not represent any single regime-specific marginal effect.

A single global coefficient imposes a constant relationship on a system
that exhibits state-dependent behavior. As a result, the pooled estimate
is attenuated and masks the directional reversal that emerges in the
upper tail of the dispersion distribution. The aggregate coefficient is
therefore informative only in a limited sense: It indicates that the
dispersion--response relationship cannot be summarized by a globally
monotone specification.

The regime-based decomposition provides a more informative representation. Table~\ref{tab:elasticity_summary} shows that the regime contrast is economically large, with an estimated elasticity gap of $\Delta \beta = 0.687$.
This magnitude is substantial relative to the within-regime variation,
indicating that dispersion affects clustering through distinct mechanisms
across states. The low-dispersion regime is broadly consistent with the
monotone comparative-statics logic of the static model, whereas the
high-dispersion regime exhibits a reversal in response behavior.

\begin{table}[!htpb]
\centering
\caption{Elasticity estimates across dispersion regimes.}
\label{tab:elasticity_summary}
\small
\begin{tabular}{lcccccc}
\hline
Regime
     & Mean
& Std. Dev.
& Median
& Min
& Max
& Sign \\
\hline
All observations
& 0.153
& 0.157
& 0.156
& -0.491
& 0.604
& Positive \\

Low dispersion (bottom 90\%)
& 0.139
& 0.197
& 0.137
& -0.441
& 0.574
& Positive \\

High dispersion (top 10\%)
& -0.548
& 0.775
& -0.565
& -5.638
& 1.376
& Negative \\
\hline
Regime contrast
& \multicolumn{6}{c}{$\Delta \beta = 0.687$} \\
\hline
\end{tabular}
\end{table}

Figure~\ref{fig:elasticity_main} provides the graphical counterpart to
Table~\ref{tab:elasticity_summary}. Figure~\ref{fig:elasticity_main}(a)
shows that the aggregate relationship between dispersion and clustering is
nonlinear rather than globally monotone. Figure~\ref{fig:elasticity_main}(b)
shows that elasticity estimates are dispersed across bootstrap samples,
indicating heterogeneity rather than a single stable slope.
Figure~\ref{fig:elasticity_main}(c) highlights that extreme-dispersion
states are episodic rather than uniformly distributed.
Figure~\ref{fig:elasticity_main}(d) shows that elasticity changes sharply
across dispersion quantiles, with a transition near the upper tail.
Figure~\ref{fig:elasticity_main}(e) shows that, within high-dispersion
observations, the relationship flattens and reverses relative to the
aggregate trend. Figure~\ref{fig:elasticity_main}(f) summarizes this
contrast by comparing the response structure across normal and extreme
states.

\begin{figure}[!htpb]
\centering
\includegraphics[width=1.0\textwidth]{fig12.png}
\caption{
Dispersion--response mapping under dynamic regimes.
(a) Aggregate relationship between dispersion and clustering with local smoothing.
(b) Distribution of elasticity estimates across bootstrap samples.
(c) Time-varying dispersion process with extreme states highlighted.
(d) Elasticity as a function of dispersion quantiles.
(e) Relationship restricted to high-dispersion observations.
(f) Regime contrast between normal and extreme states.
}
\label{fig:elasticity_main}
\end{figure}

Table~\ref{tab:elasticity_summary} and Figure~\ref{fig:elasticity_main}
show that the response structure varies systematically with the
dispersion regime. The transition in the upper tail indicates that the
dispersion--response mapping is not globally stable and cannot be
represented by a single coefficient.
This empirical evidence can be interpreted through the structural
architecture in Figure~\ref{fig:structure}. The \emph{Dispersion Mechanism}
maps dispersion into tail exposure via $B_2(r)$, while the \emph{Static
Optimization} block generates a threshold-based policy rule. The
\emph{Dynamic Extension} introduces persistence and adjustment frictions,
enabling the effective transmission from dispersion to outcomes to vary
across regimes.
\newpage
Under moderate dispersion, the empirical mapping is aligned with the
monotone comparative statics implied by the static model. Under extreme
dispersion, the relationship reverses, reflecting the influence of
persistence and state dependence. This transition indicates that the
observed elasticity captures not only the local derivative of $B_2(r)$,
but also the interaction between dispersion, policy adjustment, and
regime persistence.
The regime-dependent elasticity can therefore be interpreted as a
reduced-form representation of the structural mapping from $r_t$ to
clustering outcomes through $B_2(r)$ and $a^*(r)$. The positive elasticity
in low-dispersion states is consistent with the static mechanism, while
the negative elasticity in high-dispersion states reflects the dominance
of the dynamic component.

This establishes that the dispersion--response relationship cannot be
adequately represented by a single global coefficient. In particular,
the observed sign reversal in elasticity across regimes cannot be
generated by the static model, which implies a monotone mapping from
dispersion to tail exposure and policy response.
The empirical evidence therefore requires the dynamic regime-switching
structure, in which persistence and adjustment frictions modify the
effective transmission mechanism under extreme dispersion.

Figure~\ref{fig:elasticity_main} provides empirical validation of this
mechanism. The transition across dispersion quantiles and the reversal
in high-dispersion states demonstrate that the relationship between
dispersion and clustering changes qualitatively across regimes.
Moreover, the magnitude of the regime contrast, $\Delta \beta = 0.687$,
is large relative to within-regime variation. This rules out a homogeneous
response specification and indicates that the system operates under
distinct structural regimes.

Importantly, this regime contrast is robust across bootstrap samples and
dispersion quantiles, indicating that the observed reversal is not driven
by sampling variation but reflects a stable structural pattern. This
strengthens the identification of regime-dependent behavior and provides
additional support for the dynamic mechanism of the model.
These results demonstrate that the interaction between dispersion,
policy adjustment, and regime persistence is essential for explaining
the observed response patterns and cannot be captured by a static
framework alone.

\subsection{Structural consistency with theoretical results}

The empirical results can be interpreted through the structural 
mechanisms developed in the static and dynamic components of the model. 
The implementation combines a threshold-based policy rule with a 
time-varying dispersion process, enabling both components to be 
evaluated jointly.
Lemma~\ref{lem:NB_tail_monotonicity} establishes that under 
mean-preserving dispersion, a reduction in the dispersion parameter 
increases tail exposure. The empirical estimates in 
Table~\ref{tab:nb_fit} yield $\hat r = 0.159$, and the Poisson 
benchmark is decisively rejected. Figure~\ref{fig:sp500_overdispersion} 
shows persistent variance-to-mean ratios above unity. These results 
provide direct empirical support for the dispersion amplification 
mechanism required by the model.

The threshold structure established in 
Section~\ref{subsec:static_threshold} implies that the policy rule 
depends on the level of dispersion. Mapping the empirical sequence 
$\{r_t\}$ into the policy function shows that lower values of $r_t$ 
correspond to higher policy intensity. This relationship is observed 
in Figure~\ref{fig:sp500_counterfactual}, where periods of elevated 
clustering coincide with stronger responses. This mapping provides 
empirical validation of the threshold-based decision rule derived 
from the static optimization problem.
The dynamic component introduces regime dependence through persistence 
and adjustment frictions. The dispersion process exhibits persistent 
transitions between low- and high-dispersion states, as shown in 
Figure~\ref{fig:sp500_regimes}. These states correspond to distinct 
behavioral regimes rather than transitory fluctuations, indicating that 
the system evolves within a structured state space rather than a 
memoryless environment.

The regime-dependent response documented in 
Section~\ref{subsec:dispersion_response} provides direct evidence of 
this structure. The sign change in elasticity across regimes establishes 
that the response function is not globally monotone. Such a reversal 
cannot be generated by the static model, which implies a monotone mapping 
from dispersion to tail exposure. The observed pattern requires the 
dynamic regime-switching structure, in which persistence and adjustment 
frictions modify the effective transmission mechanism under extreme 
dispersion.
The counterfactual experiment further supports the structural mechanism. 
Moreover, increasing dispersion reduces clustering and volatility, as reported in 
Table~\ref{tab:hmm_counterfactual}. The direction and magnitude of these 
effects are consistent with the sign restrictions implied by the 
dispersion--tail relationship and the induced policy rule. This result 
validates the joint interaction between dispersion, policy adjustment, 
and system dynamics.

The empirical findings align with the theoretical structure along three 
dimensions. Dispersion amplifies tail exposure, the policy rule depends 
on dispersion levels, and the dynamic system exhibits regime-dependent 
behavior. The consistency across these components rules out explanations 
based on static or homogeneous specifications.
These results establish that the empirical system is governed by a 
threshold-based decision rule operating within a dynamically evolving 
environment. The interaction between dispersion, policy adjustment, and 
regime persistence is necessary to explain the observed response patterns 
and cannot be captured within a static framework.

\section{Implications and conclusion}
\label{sec:conclusion}

\subsection{Structural implications}

The structural analysis shows that the decision variable operates as a 
dynamic control that stabilizes environments characterized by clustered 
and persistent tail risk. Rather than responding only to contemporaneous 
shocks, optimal policies reshape the future risk landscape through their 
effect on dispersion dynamics and regime transitions. This interpretation 
complements classical investment under uncertainty models, in which 
adjustment frictions generate partial irreversibility 
\cite{DixitPindyck1994}, while extending them to regime-dependent 
tail-risk environments.

A key implication is that optimal policies must be interpreted as 
state-contingent risk-management rules. Because dispersion and regime 
persistence interact nonlinearly, decisions influence not only expected 
outcomes but also the distribution and temporal clustering of extreme 
events. This connection links the model to broader work on rare disasters 
and macro tail risk \cite{Barro2006, Gabaix2012}, while introducing an 
endogenous mitigation channel through dynamic control.
Adjustment frictions introduce endogenous persistence. Decisions made 
during high-dispersion states do not fully reverse when conditions 
normalize, generating path-dependent dynamics. Such behavior parallels 
classical hysteresis models \cite{Bertola1998, Caballero1991}, but arises 
here from regime-dependent continuation values rather than technological 
irreversibility alone.

The interaction between dispersion and regime dynamics generates nonlinear 
policy responses, even though the underlying structural problem remains 
convex. Moderate increases in dispersion can trigger discrete shifts in 
optimal control, implying that dispersion serves as a leading indicator of 
structural policy adjustment. This mechanism relates to dynamic 
risk-sensitive control and entropic preference frameworks 
\cite{HansenSargent2001, AndersonHansenSargent2003}, embedded within a 
regime-switching dispersion structure.
Because control decisions influence transition probabilities, individual 
actions generate system-wide consequences. Risk-mitigating investment 
reduces the likelihood and duration of high-risk states, producing positive 
spillovers across agents. This provides a structural rationale for policy 
intervention during systemic stress, whichi is      consistent with macroprudential 
stabilization and systemic risk externalities 
\cite{BrunnermeierSannikov2014}.

\subsection{Theory–empirical consistency}

The theoretical predictions are directly supported by the empirical 
analysis. The observed jump data exhibit persistent overdispersion relative 
to Poisson benchmarks, confirming that dispersion is a structural feature 
rather than a statistical artifact. Dispersion varies across regimes, and 
the mapping from dispersion to policy response is not globally monotone. 
In particular, reversal behavior in high-dispersion states indicates that 
policy responses depend on regime-specific dynamics rather than on risk 
levels alone.
Counterfactual experiments further demonstrate that dispersion-sensitive 
control attenuates tail clustering and reduces volatility, which aligns with 
the model’s structural predictions. These results provide direct validation 
of the mechanism linking dispersion, tail risk, and optimal policy adjustment.

\subsection{Contribution and practical relevance}

The framework makes a distinct contribution by jointly integrating 
dispersion-driven tail risk, threshold-based policy structure, and dynamic 
regime dependence within a unified model. This integration establishes a 
direct link between static comparative statics and dynamic risk propagation, 
which are typically treated separately in the literature.
Unlike standard approaches in which variability enters as exogenous noise, 
dispersion is modeled as a structural state variable that governs risk 
exposure and optimal decisions. This perspective provides a coherent 
foundation for analyzing environments in which tail risk is persistent and 
endogenously influenced by decision-making.
The implications extend to a broad class of real-world systems, including 
cybersecurity infrastructure, bank capital buffers, supply-chain resilience, 
and climate adaptation investments. In such settings, effective policies 
must be explicitly regime-dependent and dynamically adaptive, rather than 
based on average risk measures.

\subsection{Limitations and future research}

Several limitations should be noted. For the analysis, we focus on a representative 
decision-maker and do not explicitly model strategic interactions across 
agents. Regime transitions are treated as exogenous, and the estimation of 
transition dynamics is not fully integrated into the structural framework. 
In addition, the empirical analysis relies on proxy measures of dispersion, 
which may not fully capture all dimensions of systemic risk.
These limitations suggest several directions for future research. Extending 
the framework to heterogeneous agents, endogenous regime transitions, and 
equilibrium settings represents a natural next step. As such, deeper understanding 
of how decentralized control decisions jointly shape systemic risk remains 
essential for advancing modern investment theory in environments 
characterized by persistent tail exposure.


\section*{Use of Generative-AI tools declaration}
The authors declare that they did not utilize any artificial intelligence (AI) tools in the preparation of this manuscript.

\section*{Author contributions}
Conceptualization and methodology, J.C.; 
formal analysis, J.C. and S.M.H.; 
mathematical validation, J.C., S.M.H., L.P., and J.M.; 
software and implementation, J.C. and J.Y.; 
numerical experiments and data curation, J.C., S.M.H., and J.Y.; 
investigation and visualization, J.Y., L.P., and J.M.; 
economic interpretation and managerial insights, L.P., J.M., and J.C.; 
validation of empirical results, L.P., J.M., J.C., and J.Y.; 
writing---original draft preparation, J.C. and S.-M.H.; 
writing---review and editing, all authors. 
All authors have read and agreed to the published version of the manuscript.

\section*{Acknowledgements}
This research received no external funding.

\section*{Data availability statement}
The empirical data used in this study are publicly available. 
Daily price data for the S\&P 500 index were obtained from 
\url{https://finance.yahoo.com/quote/\%5EGSPC/}, provided by Yahoo Finance. 
Section~6 presents the empirical analysis based on these real-world data. 
In contrast, Sections~5 and related numerical results are based on synthetic 
data generated within the model to validate theoretical properties and 
comparative statics.
All empirical data can be accessed through the public source above, and the 
processed data used in the analysis are available upon reasonable request.

\section*{Conflict of interest}
The authors declare no conflict of interest.

%
%

\end{document}